\documentclass{bmcart}

\usepackage[utf8]{inputenc} %unicode support
\usepackage{graphicx}
\usepackage{siunitx}
    \DeclareSIUnit{\year}{yr}
    \DeclareSIUnit{\eu}{eu}
    \DeclareSIUnit{\nT}{\nano\tesla}
\usepackage{booktabs}
\usepackage{multirow}
\usepackage{amsmath}
\usepackage{amssymb}
\usepackage{array}
\usepackage{dsfont}
\usepackage{mathtools}
    \DeclarePairedDelimiter{\abs}{\lvert}{\rvert}%
    \DeclarePairedDelimiter{\diag}{\mathrm{diag}(}{)}%
    \def\doubleunderline#1{\underline{\underline{#1}}}
\usepackage{upgreek}
\usepackage[authoryear]{natbib}
\usepackage{footnote}
    \makesavenoteenv{tabular}
    \makesavenoteenv{table}
\usepackage[hyperfootnotes=false, bookmarksopen=true]{hyperref}

\newlength{\figurewidth}
\startlocaldefs
\endlocaldefs

\begin{document}

%%% Start of article front matter
\begin{frontmatter}

\begin{fmbox}
\dochead{Research}


\title{Co-estimating geomagnetic field and calibration parameters: modeling Earth's magnetic field with platform magnetometer data}

%%%%%%%%%%%%%%%%%%%%%%%%%%%%%%%%%%%%%%%%%%%%%%
%%                                          %%
%% Enter the authors here                   %%
%%                                          %%
%% Specify information, if available,       %%
%% in the form:                             %%
%%   <key>={<id1>,<id2>}                    %%
%%   <key>=                                 %%
%% Comment or delete the keys which are     %%
%% not used. Repeat \author command as much %%
%% as required.                             %%
%%                                          %%
%%%%%%%%%%%%%%%%%%%%%%%%%%%%%%%%%%%%%%%%%%%%%%

\author[
   addressref={aff1},                   % id's of addresses, e.g. {aff1,aff2}
   corref={aff1},                       % id of corresponding address, if any
   %noteref={n1},                       % id's of article notes, if any
   email={ancklo@space.dtu.dk}          % email address
]{\inits{C}\fnm{Clemens} \snm{Kloss}}
%  (blank line causes a line break in author line)
\author[
   addressref={aff1},
   email={cfinl@space.dtu.dk}
]{\inits{C. C.}\fnm{Christopher C.} \snm{Finlay}}
%  (blank line causes a line break in author line)
\author[
   addressref={aff1},
   email={nio@space.dtu.dk}
]{\inits{N}\fnm{Nils} \snm{Olsen}}

%%%%%%%%%%%%%%%%%%%%%%%%%%%%%%%%%%%%%%%%%%%%%%
%%                                          %%
%% Enter the authors' addresses here        %%
%%                                          %%
%% Repeat \address commands as much as      %%
%% required.                                %%
%%                                          %%
%%%%%%%%%%%%%%%%%%%%%%%%%%%%%%%%%%%%%%%%%%%%%%

\address[id=aff1]{  % unique id
    \orgname{Division of Geomagnetism, DTU Space, Technical University of Denmark},  % university, etc
    \street{Centrifugevej 356},
    \postcode{2800}  % post or zip code
    \city{Kongens Lyngby},  % city
    \cny{Denmark}  % country
}

%%%%%%%%%%%%%%%%%%%%%%%%%%%%%%%%%%%%%%%%%%%%%%
%%                                          %%
%% Enter short notes here                   %%
%%                                          %%
%% Short notes will be after addresses      %%
%% on first page.                           %%
%%                                          %%
%%%%%%%%%%%%%%%%%%%%%%%%%%%%%%%%%%%%%%%%%%%%%%

\begin{artnotes}
%\note{Sample of title note}     % note to the article
%\note[id=n1]{Equal contributor} % note, connected to author
\end{artnotes}

\end{fmbox}% comment this for two column layout

%%%%%%%%%%%%%%%%%%%%%%%%%%%%%%%%%%%%%%%%%%%%%%
%%                                          %%
%% The Abstract begins here                 %%
%%                                          %%
%% Please refer to the Instructions for     %%
%% authors on http://www.biomedcentral.com  %%
%% and include the section headings         %%
%% accordingly for your article type.       %%
%%                                          %%
%%%%%%%%%%%%%%%%%%%%%%%%%%%%%%%%%%%%%%%%%%%%%%

\begin{abstractbox}

\begin{abstract} % abstract

Models of the geomagnetic field rely on magnetic data of high spatial and temporal resolution to give an accurate picture of the Earth's internal magnetic field and its time-dependence. The magnetic data from low-Earth orbit satellites of dedicated magnetic survey missions such as CHAMP and \textit{Swarm} play a key role in the construction of such models. Unfortunately, there are no magnetic data available from such satellites after the end of the CHAMP mission in 2010 and before the launch of the \textit{Swarm} mission in late 2013. This limits our ability to recover signals on timescales of 3 years and less during this gap period. The magnetic data from platform magnetometers carried by satellites for navigational purposes may help address this data gap provided that they are carefully calibrated.

Earlier studies have demonstrated that platform magnetometer data can be calibrated using a fixed geomagnetic field model as reference. However, this approach can lead to biased calibration parameters. An alternative approach has been developed in the form of a co-estimation scheme which consists of simultaneously estimating both the calibration parameters and a model of the internal part of the geomagnetic field.

Here, we go further and develop a scheme, based on the CHAOS field modeling framework, that involves co-estimation of both internal and external geomagnetic field models along with calibration parameters of platform magnetometer data. Using our implementation, we are able to derive a geomagnetic field model spanning 2008 to 2018 with satellite magnetic data from CHAMP, \textit{Swarm}, secular variation data from ground observatories, and platform magnetometer data from CryoSat-2 and the GRACE satellite pair. Through a number of experiments, we explore correlations between the estimates of the geomagnetic field and the calibration parameters, and suggest how these may be avoided. We find evidence that platform magnetometer data provide additional information on the secular acceleration, especially in the Pacific during the gap between CHAMP and \textit{Swarm}. This study adds to the evidence that it is beneficial to use platform magnetometer data in geomagnetic field modeling.

\end{abstract}

%%%%%%%%%%%%%%%%%%%%%%%%%%%%%%%%%%%%%%%%%%%%%%
%%                                          %%
%% The keywords begin here                  %%
%%                                          %%
%% Put each keyword in separate \kwd{}.     %%
%%                                          %%
%%%%%%%%%%%%%%%%%%%%%%%%%%%%%%%%%%%%%%%%%%%%%%


% MSC classifications codes, if any
%\begin{keyword}[class=AMS]
%\kwd[Primary ]{}
%\kwd{}
%\kwd[; secondary ]{}
%\end{keyword}

\end{abstractbox}
%
% \end{fmbox}% uncomment this for twcolumn layout

\end{frontmatter}

%%%%%%%%%%%%%%%%%%%%%%%%%%%%%%%%%%%%%%%%%%%%%%
%%                                          %%
%% The Main Body begins here                %%
%%                                          %%
%% Please refer to the instructions for     %%
%% authors on:                              %%
%% http://www.biomedcentral.com/info/authors%%
%% and include the section headings         %%
%% accordingly for your article type.       %%
%%                                          %%
%% See the Results and Discussion section   %%
%% for details on how to create sub-sections%%
%%                                          %%
%% use \cite{...} to cite references        %%
%%  \cite{koon} and                         %%
%%  \cite{oreg,khar,zvai,xjon,schn,pond}    %%
%%  \nocite{smith,marg,hunn,advi,koha,mouse}%%
%%                                          %%
%%%%%%%%%%%%%%%%%%%%%%%%%%%%%%%%%%%%%%%%%%%%%%

\section{Introduction}
\label{sec:introduction}

The Earth’s magnetic field is a superposition of many sources. By far, the largest contribution comes from within the Earth at a depth of more than 3000 km. There, in the outer core, a liquid iron alloy is rapidly moving and thus advecting, stretching, and maintaining the ambient magnetic field against dissipation in a process called the Geodynamo. Earth’s core dynamics are not fully understood, but can be studied using time-dependent geomagnetic field models. Such models are constructed using measurements of the magnetic field taken at and above Earth’s surface.

The study of core processes on decadal or longer timescales requires long time-series of magnetic vector data with high spatial and temporal resolution. Along with ground-based magnetic observatories, low-earth orbit satellites from dedicated magnetic survey missions such as CHAllenging Minisatellite Payload (CHAMP, 2000\---2010) and the \textit{Swarm} trio (since 2013) provide such data. However, other than scalar data from {\O}rsted, no high-quality calibrated magnetic vector data from satellites are available between the end of the CHAMP mission in September 2010 and the launch of the \textit{Swarm} satellites in November 2013. This data gap not only cuts in two an otherwise uninterrupted time-series of high-quality magnetic satellite data since the year 2000, but also limits our ability to derive accurate core field models that resolve temporal changes of the magnetic field on timescales of a few years and less in the gap period. To address the issue, one can utilize the crude magnetometers that are carried by most satellites for navigational purposes, the so-called platform magnetometers. Although not a substitute for dedicated high-quality magnetic survey satellites, platform magnetometers can supplement ground observatory data in gaps between dedicated missions and help improve the local time data coverage of simultaneously flying high-quality magnetic survey satellites.

Satellite-based magnetic vector data need to be calibrated to remove magnetometer biases, scale factors, and non-orthogonalities between the three vector component axes \cite[]{Olsen2003}. Comparing the vector magnetometer output with a magnetic reference field allows the estimation of these calibration parameters. On dedicated survey mission satellites, the reference is a second, absolute scalar, magnetometer mounted in close proximity to the vector magnetometer and measuring the magnetic field intensity. However, non-dedicated satellites carrying platform magnetometers are typically not equipped with such scalar reference magnetometers. In this case, it is possible to use a-priori geomagnetic field models like CHAOS \cite[]{Olsen2006,Finlay2020} or the IGRF \cite[]{Thebault2015} as reference. Such an approach has been successfully used, e.g., by \cite{Olsen2020} for calibrating data from the \mbox{CryoSat-2} magnetometer, but use of a fixed reference field model is not without risks and could lead to biased calibration parameters.

An alternative venue has been explored by \cite{Alken2020}, who combined high-quality magnetic data from CHAMP and \textit{Swarm} with platform magnetometer data from \mbox{CryoSat-2} and several satellites of the Defense Meteorological Satellite Program (DMSP) to estimate a model of the internal field and the required calibration parameters for each satellite simultaneously. Ideally, such a co-estimation scheme eliminates the need for a-priori geomagnetic field models, but \cite{Alken2020} fall short by co-estimating only the internal field while still relying on a fixed model of the external field. Nevertheless, their study convincingly demonstrated that platform magnetometer data provide valuable information about the time-dependence of Earth's magnetic field.

In this study, we followed \cite{Alken2020} and developed a co-estimation strategy but within the framework of the CHAOS field model series. Our implementation differs in three important aspects. First, we estimated both the internal (core and crust) and external (magnetospheric) geomagnetic field contributions in contrast to only the internal field. This way, we avoided having to remove a fixed external field model from the satellite data prior to the model parameter estimation. Following the methodology of the CHAOS model, we did use a prior external field model for processing the ground observatory data which we used in addition to the satellite data. Second, we used the platform magnetometer data from \mbox{CryoSat-2} and, instead of DMSP, data from the Gravity Recovery and Climate Experiment (GRACE) satellite pair. Finally, to reduce the significant correlation between the internal axial dipole and the calibration parameters during periods of poor coverage of high-quality magnetic data, we excluded platform magnetometer data from determining the internal axial dipole (its time variation is well resolved with ground observatory data during the gap period, while its absolute value is constrained by Swarm and CHAMP data on both sides of the gap) rather than controlling the temporal variability of the internal axial dipole through an additional regularization as done by \cite{Alken2020}.

The paper is organized as follows. In the first part, we present the datasets and the data processing. Next, we describe the model parameterization and define the calibration parameters, which are similar to those used for the {\O}rsted satellite \cite[]{Olsen2003}. We go on by presenting a geomagnetic field model derived from high-quality calibrated data from the CHAMP and the \textit{Swarm} satellites as well as ground observatory secular variation data and supplemented this with previously uncalibrated platform magnetometer data from \mbox{CryoSat-2} and GRACE, spanning a 10 year period from 2008 to 2018. Finally, we explore in a series of experiments the effect of co-estimating an external field, the trade-off between the internal dipole and the calibration parameters, and the importance of including dayside platform magnetometer data when estimating calibration parameters. We conclude the paper by looking at the secular acceleration of our model, paying particular attention to the data gap between 2010 and 2013.

\section{Data and data processing}
\label{sec:data_and_data_processing}

We used calibrated magnetic data from the \textit{Swarm} satellites Alpha (\mbox{Swarm-A}) and Bravo (\mbox{Swarm-B}), and from the CHAMP satellite from January 2008 to the end of December 2017, supplemented with five datasets of uncalibrated magnetic data from the three platform fluxgate magnetometers (FGM) on-board the \mbox{CryoSat-2} satellite (\mbox{CryoSat-2} FGM1, \mbox{CryoSat-2} FGM2 and \mbox{CryoSat-2} FGM3), the one on-board the first GRACE satellite (\mbox{GRACE-A}), and the other one on-board the second GRACE satellite (\mbox{GRACE-B}). In addition to the satellite data, we included revised monthly mean values of the SV from ground observatories to contribute to the Earth's internal time-dependent field. Details of the datasets are given in the following.

\subsection{Absolute satellite data from scientific magnetometers}

The satellite data from scientific magnetometers are in general of high quality in terms of accuracy, precision and magnetic cleanliness. The high standard of the data is achieved by low noise instruments that are mounted together with star cameras on an optical bench further away from the spacecraft body at the center of a several meter long boom. The data are regularly calibrated in-flight with a second absolute scalar magnetometer placed at the end of the boom and carefully cleaned from magnetic disturbance fields originating from the spacecraft body.

From the CHAMP mission, we used the Level 3 \SI{1}{\hertz} magnetic data, version CH-ME-3-MAG \cite[]{Rother2019}, between January 2008 and August 2010, downsampled to \SI{15}{\second}, and only when attitude information from both star cameras was available. From the \textit{Swarm} mission, we used the Level 1b \SI{1}{\hertz} magnetic data product, baseline 0505/0506, from the \mbox{Swarm-A} and \mbox{Swarm-B} satellites between November 2013 and December 2018, also downsampled to \SI{15}{\second}. Here, we worked with vector data from CHAMP and \textit{Swarm} in the magnetometer frame.

\subsection{Relative satellite data from platform magnetometers}

Relative satellite data refer to the raw sensor output from platform magnetometers. The data have to be corrected and calibrated before they can be used in geomagnetic field modeling. The correction of the data accounts for temperature effects, magnetic disturbances due to solar array and battery currents, magnetorquer activity, as well as non-linear sensor effects, whereas the calibration removes magnetometer biases, scale differences, and non-orthogonalities between the three vector component axes.

\begin{figure*}
\centering
\includegraphics[width=\figurewidth]{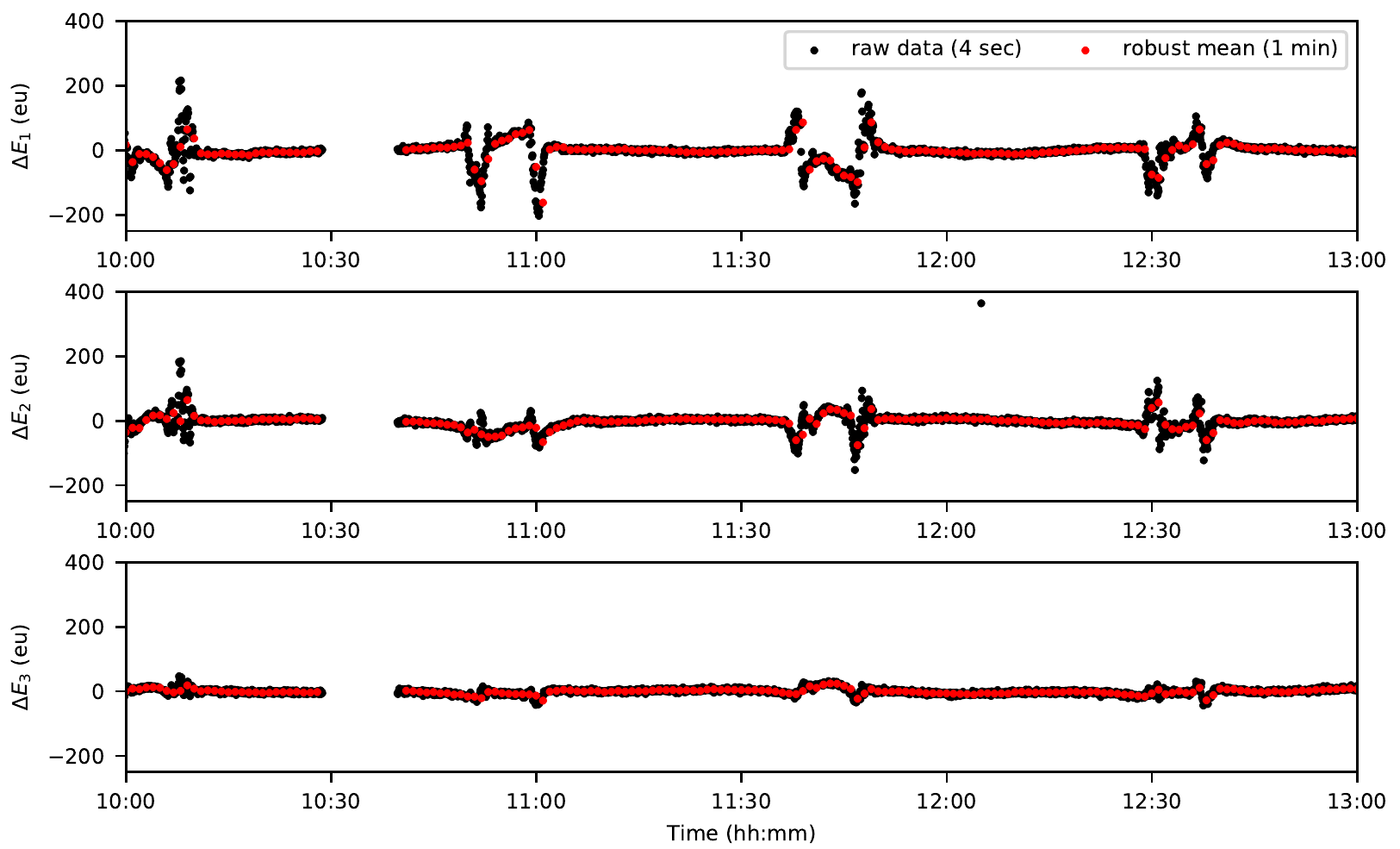}
\caption{Residuals of raw vector data from CryoSat-2 FGM1 with respect to the CHAOS-6-x9 model values in the uncalibrated magnetometer frame for an example period of \SI{3}{\hour} on March 24, 2016. The gap in the raw data between 10:30 and 10:40 is due to the rejection of data with poor attitude information ($q_\mathrm{error} > \SI{40}{\arcsecond}$).}
\label{fig:comparison_raw_data}
\end{figure*}

From \mbox{CryoSat-2}, we took magnetic data, baseline 0103, from the three platform magnetometers as described in \cite{Olsen2020} from August 2010 to December 2018 and only when the attitude uncertainty $q_\mathrm{error}$ was below \SI{40}{\arcsecond}. Since the purpose of this paper is the co-estimation of calibration parameters for the platform magnetometers, we processed the dataset using the original calibration parameters to undo the calibration step that has been performed by \cite{Olsen2020} but keeping the applied correction for magnetic disturbances from the spacecraft and its payload. This way, we obtained essentially uncalibrated data while still retaining the corrections for magnetic disturbances, temperature effects and non-linearities. In a pre-whitening and data reduction step, we computed residuals to the \mbox{CHAOS-6-x9} model in the uncalibrated magnetometer frame, removed those larger than \SI{1000}{\eu} (quasi nanoTesla, in the following referred to as engineering units) in absolute value to discard gross outliers, computed component-wise robust mean values of the residuals in \SI{1}{min} bins to reduce the original \SI{4}{\second} sampled data to \SI{1}{min} values, and added the \mbox{CHAOS-6-x9} model values back. Fig.~\ref{fig:comparison_raw_data} shows an example of the raw vector residuals $\Delta\mathbf{E}$ of CryoSat-2 FGM1 in the uncalibrated magnetometer frame over \SI{3}{\hour} on March 24, 2016.
In a similar way, we processed the \SI{1}{\hertz} data from the GRACE satellites, baseline 0101, to obtain \SI{1}{min} uncalibrated but corrected vector data between January 2008 and October 2017 (\mbox{GRACE-A}) and August 2017 (\mbox{GRACE-B}) \cite[]{Olsen2020a}.

The computation of \SI{1}{\minute} values served two purposes. First, to reduce the random noise of the magnetometers by taking the average of successive values and, second, to decrease the number of platform magnetometer data, so that a fair amount of absolute satellite data was able to guide the co-estimation of the calibration parameters.

\subsection{Ground observatory data}

In addition to satellite data, we added annual differences of monthly mean values from 162 ground observatories to help determine the time changes of the core field (secular variation). Following \cite{Olsen2014}, we computed revised monthly means as Huber-weighted averages of the hourly observatory mean values from the AUX OBS database \cite[]{Macmillan2013} at all local times after removing estimates of the ionospheric field of the CM4 model \cite[]{Sabaka2004} and the large-scale magnetospheric field of \mbox{CHAOS-6-x9}, including their internally induced parts.

\subsection{Satellite data selection}

We organized the satellite data according to quasi-dipole (QD) latitude \cite[]{Richmond1995} into a non-polar (equal to and equatorward of \ang{\pm 55}) and a polar (poleward of \ang{\pm 55}) data subset. From each subset, we selected data under quiet geomagnetic conditions. Specifically, we selected data from the non-polar subset that satisfied the following criteria:
\begin{itemize}
    \item Low geomagnetic activity as indicated by the planetary activity index Kp smaller than or equal to $2^o$;
    \item Dark condition as indicated by a solar zenith angle greater than \ang{100} for the \textit{Swarm} and CHAMP satellites (i.e., sun at least \ang{10} below the horizon). From \mbox{CryoSat-2} and GRACE, we used data from dark and sunlit regions, since we found that this leads to better determined calibration parameters;
    \item Slow change of the magnetospheric ring current as indicated by the RC-index \cite[]{Olsen2014} rate of change in absolute terms being smaller than \SI{2}{\nT\per\hour}.
\end{itemize}
From the polar subset, we kept data according to the following criteria:
\begin{itemize}
    \item Dark condition except in the case of platform magnetometers on-board \mbox{CryoSat-2} and GRACE, where we also used sunlit data;
    \item RC-index rate of change in absolute terms smaller than or equal to \SI{2}{\nT\per\hour};
    \item The merging electric field at the magnetopause $E_\mathrm{m}=v^{4/3}B_\mathrm{T}^{2/3}\sin\abs{\Theta}/2$, where $v$ is the solar wind speed, $B_\mathrm{T}=\sqrt{B_y^2+B_z^2}$ is the interplanetary magnetic field in the $y$\---$z$-plane of the Geocentric Solar Magnetic (GSM) coordinates, and $\Theta=\arctan(B_y/B_z)$, was on average smaller than \SI{2.4}{\milli\volt\per\meter} over the previous \SI{2}{\hour};
    \item The interplanetary magnetic field component $B_z$ in GSM coordinates was on average positive over the previous \SI{2}{\hour}.
\end{itemize}

\begin{figure*}
    \includegraphics[width=\figurewidth]{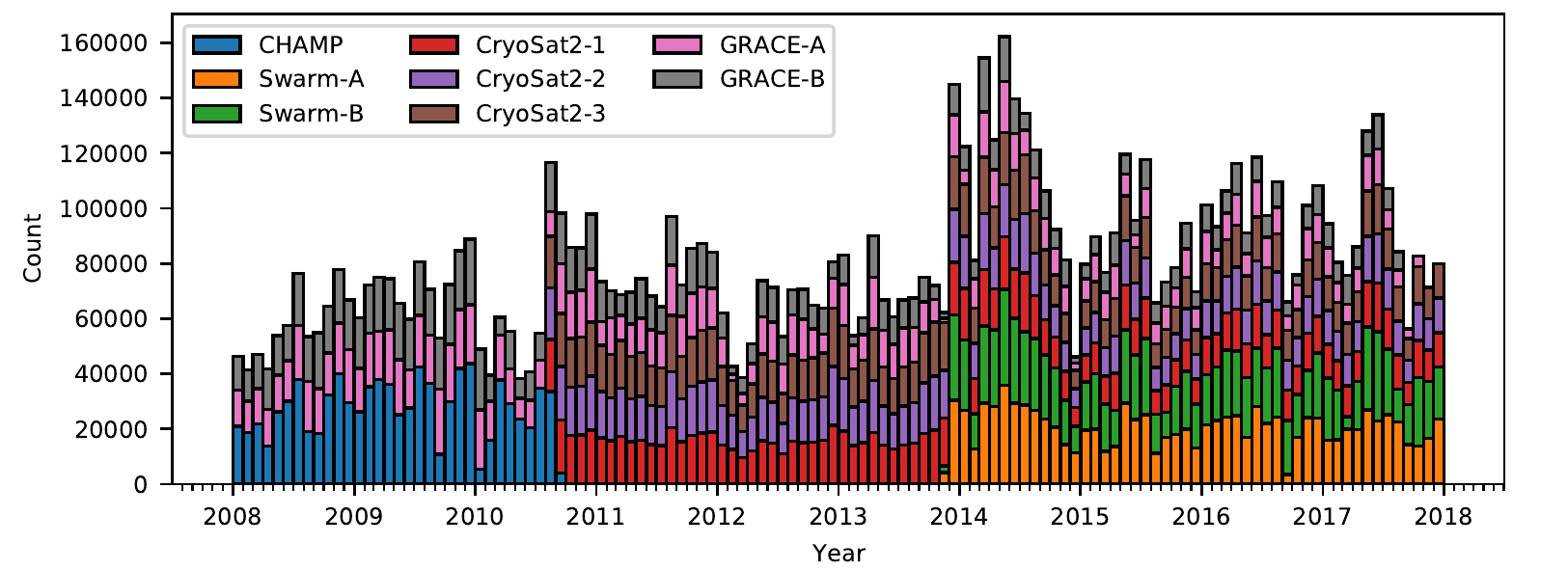}
    \caption{Number of selected scalar and vector satellite data per month as stacked histogram. Ground observatories contribute with approximately 130 vector measurements of the SV per month.}
    \label{fig:data_distribution}
\end{figure*}

Fig.~\ref{fig:data_distribution} shows a stacked histogram of the number of data for each satellite after the data selection.
It can be clearly seen that platform magnetometer data are the main contributor to the number of data in the gap period, whereas it is comparable to the number of data from CHAMP and the \textit{Swarm} satellites in the time before and after the gap. The ground observatories contribute approximately 130 monthly mean values of the SV each month throughout the entire model time span, which is much less than the monthly average number of satellite data.

\section{Model parameterization and estimation}
\label{sec:methods}

We are interested in the magnetic field vector $\mathbf{B}$ on length scales smaller than Earth's circumference and time scales that are much longer than the time it takes light to traverse these distances \citep{Backus1996,Sabaka2010}. On these scales, the displacement current can be neglected and the magnetic field is governed by Ampere's law. We assume that the measurements of Earth's magnetic field are taken in a region free of electrical currents and magnetized material, such that the field is irrotational, which allows us to introduce a scalar potential $V$ to represent the magnetic field as the gradient of the potential $\mathbf{B} = -\nabla V$. The potential consists of two terms $V = V_\mathrm{int} + V_\mathrm{ext}$ that describe internal sources such as the time-dependent core-generated field and the assumed static lithospheric field, and external sources that we assume are mainly magnetospheric in origin for our chosen data selection criteria and have an internally induced counterpart associated with them (by selecting data from dark regions, we minimize ionospheric field contributions).

To describe the geomagnetic field, we use an Earth-fixed frame of reference whose point of origin coincides with the Earth's center and in which the position vector $\mathbf{r}$ is given in spherical polar coordinates by the radial distance $r$ as measured from the origin (radius), the angular distance $\theta$ (co-latitude) as measured from the north polar axis, and the azimuthal angular distance $\phi$ (longitude) as measured from the Greenwich meridian. In the following, we refer to that system as the Radius-Theta-Phi (RTP) reference frame.

In spherical coordinates the scalar potential can be expressed as a weighted sum of solid harmonics, which are harmonic functions of the spatial coordinates. Our modeling approach follows that of earlier models of the CHAOS model series \cite[]{Olsen2006,Olsen2014,Finlay2016, Finlay2020} and consists of describing the geomagnetic field with the help of a scalar potential whose exact form depends on a set of coefficients that multiply the solid harmonics. The coefficients are estimated by minimizing a quadratic cost function in the residuals, the difference between the magnetic observations and the magnetic data calculated with the model. We used two kinds of residuals: the components of vector differences in the RTP frame (vector residuals) and the difference of vector magnitudes (scalar residuals). More specifically, we computed vector residuals of the non-polar satellite data, scalar residuals of the polar satellite data, and vector residuals of the ground observatory SV data at all QD latitudes.

\subsection{Internal field parameters}
\label{sec:internal_field_parameters}

The scalar potential of the internal sources is given by
\begin{equation}
    \label{eq:potential_int}
    V_\mathrm{int}(\mathbf{r}, t) = a\sum_{n=1}^{N_\mathrm{int}} \sum_{m=0}^n (g_n^m(t) \cos m\phi + h_n^m(t) \sin m\phi)\left(\frac{a}{r}\right)^{n+1} P_n^m(\cos\theta),
\end{equation}
where $a = \SI{6371.2}{\kilo\meter}$ is the chosen spherical reference radius of the Earth, $n$ and  $m$ are, respectively, the spherical harmonic degree and order, $N_\mathrm{int}$ is the truncation degree, $g_n^m(t)$ and $h_n^m(t)$ are the Gauss coefficients in nanoTesla (\si{\nT}) for a given $n$ and $m$, and $P_n^m(\cos\theta)$ are the Schmidt quasi-normalized associated Legendre functions. We truncated the formally infinite sum of solid harmonics at $N_\mathrm{int} = 50$ and expanded the Gauss coefficients of degree $n \leq 15$ in time using sixth-order B-splines \cite[]{DeBoor1978}, while we kept the higher degree coefficients ($n > 15$) constant in time
\begin{equation}
g_n^m(t) = \left\{
    \begin{aligned}
    &\sum_{j}g_{n,j}^m B_{6,j}(t), \quad n\leq 15\\
    & g_n^m, \quad n > 15,
    \end{aligned}
\right.
\end{equation}
where $g_{n,j}^m$ (similarly for $h_{n,j}^m$) is the coefficient of $B_{6,j}(t)$\----the $j$th function of the B-spline basis that has knots at 6-month intervals and six-fold multiplicity at the model endpoints in $t_\mathrm{s}=\num{2008.0}$ and $t_\mathrm{e}=\num{2018.0}$ in years. For the purposes of testing the co-estimation of calibration parameters here, a truncation of the time-dependent internal field at degree $N_\mathrm{int}=15$ was deemed sufficient.

\subsection{External field parameters}
\label{sec:external_field_parameters}

The scalar potential of the external sources consists of two terms $V_\mathrm{ext} = V_\mathrm{SM} + V_\mathrm{GSM}$ that are designed to account for near and remote magnetospheric sources. We use the Solar Magnetic (SM) coordinate system to parameterize near magnetospheric sources
\begin{equation}
    \label{eq:potential_sm}
    \begin{aligned}
        V_\mathrm{SM} &= a\sum_{m=0}^{1}\big(q_{1,\mathrm{SM}}^{m} (r,t) \cos m\phi_\mathrm{SM} + s_{1, \mathrm{SM}}^{m} (r,t) \sin m\phi_\mathrm{SM}\big) P^m_1(\cos \theta_\mathrm{SM})\\
        &+ a\sum_{m=0}^{1}\big(\Delta q_{1, \mathrm{SM}}^{m}(t)R_{1,\mathrm{SM}}^{m,\mathrm{c}}(\mathbf{r}, t) + \Delta s_{1, \mathrm{SM}}^{m}(t)R_{1,\mathrm{SM}}^{m,\mathrm{s}}(\mathbf{r}, t)\big)\\
        &+ a\sum_{m=0}^{2}\big(q_{2,\mathrm{SM}}^{m,\mathrm{c}}R_{2,\mathrm{SM}}^{m,\mathrm{c}}(\mathbf{r}, t) + s_{2, \mathrm{SM}}^m R_{2,\mathrm{SM}}^{m,\mathrm{s}}(\mathbf{r}, t)\big),
    \end{aligned}
\end{equation}
where $\theta_\mathrm{SM}$ and $\phi_\mathrm{SM}$ are, respectively, the SM co-latitude and longitude, $q_{n,\mathrm{SM}}^{m}$ and $s_{n,\mathrm{SM}}^{m}$ are the Gauss coefficients with respect to the SM coordinate system, $\Delta q_{1, \mathrm{SM}}^{m}(t)$ and $\Delta s_{1, \mathrm{SM}}^{m}(t)$ are the RC-baseline corrections, and $R_{n,\mathrm{SM}}^{m,\mathrm{s}}$ and $R_{n,\mathrm{SM}}^{m,\mathrm{c}}$ are modification of the solid harmonics that account for the time-dependent transformation from the SM to the geographic coordinate system and include internally induced contributions based on the diagonal part of the Q-response matrix that has been derived from a 3D conductivity model of Earth \cite[]{Finlay2020}. The external Gauss coefficients with $n=1$ have a specific time-dependence in the form of
\begin{equation}
    \begin{aligned}
        q_{1,\mathrm{SM}}^{0} (r,t) &= \hat{q}_1^0 \left[\epsilon(t)\left(\frac{r}{a}\right)+\iota(t)\left(\frac{a}{r}\right)^2\right]\\
        q_{1,\mathrm{SM}}^{1} (r,t) &= \hat{q}_1^1 \left[\epsilon(t)\left(\frac{r}{a}\right)+\iota(t)\left(\frac{a}{r}\right)^2\right]\\
        s_{1,\mathrm{SM}}^{1} (r,t) &= \hat{s}_1^1 \left[\epsilon(t)\left(\frac{r}{a}\right)+\iota(t)\left(\frac{a}{r}\right)^2\right],
    \end{aligned}
\end{equation}
where $\epsilon(t)$ and $\iota(t)$ are the respective internal and external part of the RC-index linearly interpolated from hourly values. The RC-baseline corrections were estimated in bins of 30 days except in the gap period, where we used a single bin from August 2010 to January 2014 to reduce the strong co-linearity between the calibration parameters and the baseline corrections that earlier tests had revealed.

The remote magnetospheric sources, the currents at the magnetopause and in the magnetotail, are taken into account by a purely zonal potential in the GSM coordinate system up to degree 2
\begin{equation}
    V_\mathrm{GSM}(\mathbf{r}, t) = a\sum_{n=1}^{2}q_{n, \mathrm{GSM}}^0 R_{n, \mathrm{GSM}}^{0, \mathrm{c}}(\mathbf{r}, t),
\end{equation}
where $q_{n,\mathrm{GSM}}^{m}$ and $s_{n,\mathrm{GSM}}^{m}$ are Gauss coefficients that are constant in time with respect to the GSM coordinate system, and $R_{n,\mathrm{GSM}}^{m,\mathrm{c}}$ are modifications of the solid harmonics similar to corresponding terms in Eq.~\eqref{eq:potential_sm} but for the GSM coordinates.

\subsection{Alignment parameters}
\label{sec:alignment_parameters}

Using satellite data in the vector field magnetometer frame (VFM) requires an additional step, called data alignment, which involves determining alignment parameters that describe the rotation of the magnetic field vector $\mathbf{B}_\mathrm{VFM}$ in the VFM frame to $\mathbf{B}_\mathrm{CRF}$ in the common reference frame (CRF) of the satellite. Once in the CRF, the vector components can be combined with the attitude information from the star camera and rotated into the RTP frame for computing the vector residuals. We performed the data alignment for CHAMP, \textit{Swarm}, \mbox{CryoSat-2}, and GRACE.

The alignment parameters are usually parameterized in the form of Euler angles $\alpha$, $\beta$, and $\gamma$. We adopted the 1-2-3 convention of the Euler angles to align the magnetic field
\begin{equation}
\label{eq:alignment}
    \begin{aligned}
        \mathbf{B}_\mathrm{CRF} &= \doubleunderline{\mathbf{R}_\mathrm{CRF}^\mathrm{VFM}}(\alpha, \beta,\gamma) \mathbf{B}_\mathrm{VFM}\\
        &= \doubleunderline{\mathbf{R}_3}(\gamma) \doubleunderline{\mathbf{R}_2}(\beta) \doubleunderline{\mathbf{R}_1}(\alpha) \mathbf{B}_\mathrm{VFM},
    \end{aligned}
\end{equation}
where the rotation matrix is a combination of the three rotations
\begin{equation}
    \begin{aligned}
        \doubleunderline{\mathbf{R}_1} &=
        \begin{pmatrix}
        1 & 0 & 0 \\
        0 & \cos\alpha & -\sin\alpha \\
        0 & \sin\alpha & \cos\alpha \\
        \end{pmatrix}
        \\
        \doubleunderline{\mathbf{R}_2} &=
        \begin{pmatrix}
        \cos\beta & 0 & \sin\beta \\
        0 & 1 & 0 \\
        -\sin\beta & 0 & \cos\beta \\
        \end{pmatrix}
        \\
        \doubleunderline{\mathbf{R}_3} &=
        \begin{pmatrix}
        \cos\gamma & -\sin\gamma & 0 \\
        \sin\gamma & \cos\gamma & 0 \\
        0 & 0 & 1 \\
        \end{pmatrix}.
    \end{aligned}
\end{equation}
Following the alignment, we applied another rotation matrix $\doubleunderline{\mathbf{R}_\mathrm{RTP}^\mathrm{CRF}}$ to rotate the field components from the CRF to the RTP reference frame
\begin{equation}
\label{eq:use_rotation}
    \mathbf{B}_\mathrm{RTP} = \doubleunderline{\mathbf{R}_\mathrm{RTP}^\mathrm{CRF}}(\mathbf{r},t)\mathbf{B}_\mathrm{CRF},
\end{equation}
which depends on position and time. That rotation matrix was computed by combining the quaternions that express the rotation from the CRF to the Earth-fixed Earth-centered North-East-Center (NEC) frame with quaternions that describe the change from the NEC to the RTP reference frame.  For each satellite dataset, we parameterized the Euler angles in time as a piecewise constant function using a sequence of 30 day bins.

\subsection{Calibration parameters}
\label{sec:calibration_parameters}

The calibration can be viewed as an extension of the data alignment which makes it possible to use platform magnetometer data in geomagnetic field modeling. We performed the calibration for CryoSat-2 and the GRACE satellites.

We assume that the platform magnetometer is a linear vector field magnetometer, which provides information about the desired local magnetic field vector $\mathbf{B}_\mathrm{VFM}$ (units of \si{\nT}) in the form of the sensor output $\mathbf{E}=(E_1,E_2,E_3)^\mathrm{T}$ (units of \si{\eu}), which typically consists of components that are measured relative to three biased and non-orthogonal axes employing different scale factors \cite[]{Olsen2003}. More specifically, the sensor output in the magnetometer frame is related to the local magnetic field through
\begin{equation}
\label{eq:calibration}
    \mathbf{B}_\mathrm{VFM} = \doubleunderline{\mathbf{P}}^{-1}\doubleunderline{\mathbf{S}}^{-1}(\mathbf{E}-\mathbf{b}),
\end{equation}
where
\begin{equation}
    \doubleunderline{\mathbf{S}}(\mathbf{s})=
    \begin{pmatrix}
        s_1 & 0 & 0 \\
        0 & s_2 & 0 \\
        0 & 0 & s_3
    \end{pmatrix}
\end{equation}
is the diagonal matrix of sensitivities or scale factors $\mathbf{s} = (s_1, s_2, s_3)^\mathrm{T}$ (units of \si{\eu\per\nT}),
\begin{equation}
    \doubleunderline{\mathbf{P}}(\mathbf{u})=
    \begin{pmatrix}
        1 & 0 & 0 \\
        -\sin u_1 & \cos u_1 & 0 \\
        \sin u_2 & \sin u_3 & \sqrt{1-\sin^2 u_2-\sin^2 u_3}
    \end{pmatrix}
\end{equation}
is the matrix that projects the orthogonal components of magnetic field vector $\mathbf{B}_\mathrm{VFM}$ onto three non-orthogonal directions defined by the non-orthogonality angles $\mathbf{u} = (u_1, u_2, u_3)^\mathrm{T}$ (no units), and
\begin{equation}
    \mathbf{b}=
    \begin{pmatrix}
        b_1 \\
        b_2 \\
        b_3
    \end{pmatrix}
\end{equation}
is the offset or bias vector (units of \si{eu}). Combining the calibration step in Eq.~\eqref{eq:calibration}, the alignment step involving the Euler angles in Eq.~\eqref{eq:alignment} and the change of frame in Eq.~\eqref{eq:use_rotation}, yields an equation that transforms the uncalibrated sensor output $\mathbf{E}$ into calibrated, aligned field components in the RTP frame
\begin{equation}
\label{eq:calibration_alignment_rotation}
    \mathbf{B}_\mathrm{RTP} = \doubleunderline{\mathbf{R}_\mathrm{RTP}^\mathrm{CRF}}(r, \theta, \phi)\doubleunderline{\mathbf{R}_\mathrm{CRF}^\mathrm{VFM}}(\alpha, \beta, \gamma)\doubleunderline{\mathbf{P}}^{-1}\doubleunderline{\mathbf{S}}^{-1}[\mathbf{E}-\mathbf{b}].
\end{equation}
We estimated the nine basic calibration parameters and the three Euler angles in bins of 30 days. For data equatorward of \ang{\pm 55} QD latitude, we performed a vector calibration using the component residuals of $\mathbf{B}_\mathrm{RTP}$ for estimating the model parameters (see Sec.~\nameref{sec:model_parameter_estimation}). In contrast, for data poleward of \ang{\pm 55} QD latitude, we performed a scalar calibration by using the residuals of the vector magnitude, in which case the rotation matrices from the VFM to the RTP frame including the Euler angles disappear
\begin{equation}
    \begin{aligned}
        F &= \abs{\mathbf{B}_\mathrm{RTP}} = \sqrt{\mathbf{B}^\mathrm{T}_\mathrm{RTP}\mathbf{B}^\mathrm{\vphantom{T}}_\mathrm{RTP}}\\
        &= \sqrt{(\mathbf{E}-\mathbf{b})^\mathrm{T}\doubleunderline{\mathbf{S}}^{-1}(\doubleunderline{\mathbf{P}}^{-1})^\mathrm{T} \doubleunderline{\mathbf{P}}^{-1}\doubleunderline{\mathbf{S}}^{-1}(\mathbf{E}-\mathbf{b})}
    \end{aligned}
\end{equation}
at the expense of loosing the ability to estimate the Euler angles.

Tab.~\ref{tab:model_details} summarizes the different parts of the model and the corresponding number of parameters.
\begin{table*}
    \caption{Details on the parameterization of the individual model parts. Here, the number of basic parameters refers to the number of parameters irrespective of an explicit time-dependence.}
    \label{tab:model_details}
    \begin{tabular}{llccc}
        \toprule
        & & Number of & Temporal & Number of\\
        & & basic parameters & parameterization & parameters \\
        \multicolumn{2}{l}{Description of the model parameters} & & & \\
        \midrule
        \multirow{2}{*}{Internal field} & Time-dependent ($n \leq 15$) & 255 & order-6 B-spline & 6375 \\
        & Static ($16 \leq n \leq 50$) & 2,345 & None & 2345 \\
        \midrule
        \multirow{3}{*}{External field} & SM degree-1 & 3 & RC-index & 3\\
        & SM degree-2 & 5 & None & 5 \\
        & RC-baseline corrections & 3 & 80 bins (30 days) & 240\\
        & GSM & 2 & None & 2 \\
        \midrule
        \multirow{3}{*}{Euler angles} & CHAMP & 3 & 33 bins (30 days) & 99\\
        & \mbox{Swarm-A} & 3 & 50 bins (30 days) & 150\\
        & \mbox{Swarm-B} & 3 & 50 bins (30 days) & 150\\
        \midrule
        \multirow{5}{*}{Euler/Calibration} & \mbox{CryoSat-2} FGM1 & 12 & 91 bins (30 days) & 1092\\
        & \mbox{CryoSat-2} FGM2 & 12 & 91 bins (30 days) & 1092\\
        & \mbox{CryoSat-2} FGM3 & 12 & 91 bins (30 days) & 1092\\
        & \mbox{GRACE-A} & 12 & 120 bins (30 days) & 1440\\
        & \mbox{GRACE-B} & 12 & 118 bins (30 days) & 1416\\
        \midrule
        \multicolumn{4}{r}{Total number of parameters (no platform magnetometer data)} & 9369\\
        \multicolumn{4}{r}{Total number of parameters} & 15501\\
        \bottomrule
    \end{tabular}
\end{table*}

\subsection{Model parameter estimation}
\label{sec:model_parameter_estimation}

The geomagnetic field model parameters $\mathbf{p}$, the Euler angles $\mathbf{q}$, and the calibration parameters $\mathbf{e}$ were derived by solving the least-squares problem
\begin{equation}
    \label{eq:optimization}
    \mathbf{m}^* = \underset{\mathbf{m}}{\mathrm{argmin}}\,\Phi(\mathbf{m}),
\end{equation}
where $\mathbf{m} = (\mathbf{p}^\mathrm{T}, \mathbf{q}^\mathrm{T}, \mathbf{e}^\mathrm{T})^\mathrm{T}$ is the entire model parameter vector, and $\Phi$ is the cost function
\begin{equation}
    \Phi(\mathbf{m}) = \big(\mathbf{g}(\mathbf{p})-\mathbf{d}(\mathbf{q}, \mathbf{e})\big)^\mathrm{T}\doubleunderline{\mathbf{C}_\mathrm{d}}^{-1}\big(\mathbf{g}(\mathbf{p})-\mathbf{d}(\mathbf{q}, \mathbf{e})\big) + \mathbf{m}^\mathrm{T}\doubleunderline{\mathbf{\Lambda}}\mathbf{m},
\end{equation}
which penalizes a quadratic form in the residuals\----the difference between the computed geomagnetic field model values $\mathbf{g}(\mathbf{p})$ and the calibrated, aligned magnetic data $\mathbf{d}(\mathbf{q}, \mathbf{e})$\----using the inverse of the data covariance matrix $\doubleunderline{\mathbf{C}_\mathrm{d}}$, and a quadratic form in the model parameter vector using the regularization matrix $\doubleunderline{\mathbf{\Lambda}}$. For the definition of the matrices $\doubleunderline{\mathbf{C}_\mathrm{d}}$ and $\doubleunderline{\mathbf{\Lambda}}$, see, respectively, Secs.~\nameref{sec:data_weighting} and \nameref{sec:model_regularization}.

The least-squares solution $\mathbf{m}^*$ in Eq.~\eqref{eq:optimization} is found through an iterative quasi-Newton method, which consists of updating the model parameter vector $\mathbf{m}_k$ at iteration $k$ using $\mathbf{m}_{k+1} = \mathbf{m}_k + \Delta\mathbf{m}$ together with
\begin{equation}
    \label{eq:iterative_procedure}
    \begin{aligned}
        \Delta \mathbf{m} = \big(&(\doubleunderline{\mathbf{G}}_k)^\mathrm{T}\doubleunderline{\mathbf{C}_\mathrm{d}}^{-1}\doubleunderline{\mathbf{G}}_k+\doubleunderline{\mathbf{\Lambda}}\big)^{-1}\\
        &\cdot\big((\doubleunderline{\mathbf{G}}_k)^\mathrm{T}\doubleunderline{\mathbf{C}_\mathrm{d}}^{-1}(\mathbf{d}_k-\mathbf{g}_k)-\doubleunderline{\mathbf{\Lambda}}\mathbf{m}_k\big),
    \end{aligned}
\end{equation}
where $\mathbf{d}_k = \mathbf{d}(\mathbf{q}_k, \mathbf{e}_k)$, $\mathbf{g}_k = \mathbf{g}(\mathbf{p}_k)$, and $\doubleunderline{\mathbf{G}}_k$ is a matrix with entries corresponding to the partial derivative of the $i$th residual with respect to the $j$th model parameter
\begin{equation}
    \label{eq:jacobian}
    \big(\doubleunderline{\mathbf{G}}_k\big)_{ij} = \frac{\partial \big(\mathbf{g}(\mathbf{p})-\mathbf{d}(\mathbf{q}, \mathbf{e})\big)_i}{\partial (\mathbf{m})_j}\bigg|_{\mathbf{m} = \mathbf{m}_k}
\end{equation}
evaluated at iteration $k$ \cite[p.~69]{Tarantola2005}. Some entries of $\doubleunderline{\mathbf{G}}_k$ are zero owing to data subsets that do not provide information on parts of the model. For example, the scalar data do not constrain the Euler angles and the vector data from one magnetometer do not constrain the Euler angles associated with another magnetometer. With the same idea in mind, we modified entries of $\doubleunderline{\mathbf{G}}_k$ to prevent some data subsets from constraining certain parts of the internal field model. In particular, we set entries to zero for the following criteria:
\begin{enumerate}
    \item The row index of the matrix entry corresponded to dayside data from a platform magnetometer, on-board \mbox{CryoSat-2} or GRACE, and the column index corresponded to model parameters that describe the internal and external magnetic field. Therefore, the dayside data were only used to constrain the Euler angles and calibration parameters of the respective platform magnetometer.
    \item The row index of the matrix entry corresponded to data from a platform magnetometer, on-board \mbox{CryoSat-2} or GRACE, and the column index corresponded to the B-spline parameters that parameterize the $g_1^0$ Gauss coefficient of the internal field in time. Therefore, no platform magnetometer data were used to constrain the B-spline coefficients of the axial dipole which we believe are well determined using ground observatory data.
\end{enumerate}
Tab.~\ref{tab:jacobian_zeros} gives an overview of whether or not certain datasets constrained specific parts of the model.
\begin{table*}
    \caption{Overview of which data subset constrained which part of the model. The cross refers to non-zero entries in the matrix of partial derivatives, whereas the circle refers to zeros. The SV data refer to the annual difference of the revised monthly means.}
    \label{tab:jacobian_zeros}
    \begin{tabular}{llccccc}
        \toprule
        & & \multicolumn{2}{c}{Non-polar satellite data} & \multicolumn{2}{c}{Polar satellite data} & SV data\\
        & & Day & Night & Day & Night & \\
        \multicolumn{2}{l}{Description of the model parameters} & & & & & \\
        \midrule
        \multirow{2}{*}{Internal field} & Time-dependent ($n \leq 15$) & $\bigcirc$ & X$^{1}$ & $\bigcirc$ & X$^{1}$ & X \\
        & Static ($16 \leq n \leq 50$) & $\bigcirc$ & X & $\bigcirc$ & X & $\bigcirc$ \\
        \midrule
        \multirow{2}{*}{External field} & SM & $\bigcirc$ & X & $\bigcirc$ & X & $\bigcirc$ \\
        & GSM & $\bigcirc$ & X & $\bigcirc$ & X & $\bigcirc$ \\
        \midrule
        \multirow{8}{*}{Euler angles} & CHAMP & $\bigcirc$ & X & $\bigcirc$ & $\bigcirc$ & $\bigcirc$ \\
        & \mbox{Swarm-A} & $\bigcirc$ & X & $\bigcirc$ & $\bigcirc$ & $\bigcirc$ \\
        & \mbox{Swarm-B} & $\bigcirc$ & X & $\bigcirc$ & $\bigcirc$ & $\bigcirc$ \\
        & \mbox{CryoSat-2} FGM1 & X & X & $\bigcirc$ & $\bigcirc$ & $\bigcirc$ \\
        & \mbox{CryoSat-2} FGM2 & X & X & $\bigcirc$ & $\bigcirc$ & $\bigcirc$ \\
        & \mbox{CryoSat-2} FGM3 & X & X & $\bigcirc$ & $\bigcirc$ & $\bigcirc$ \\
        & \mbox{GRACE-A} & X & X & $\bigcirc$ & $\bigcirc$ & $\bigcirc$ \\
        & \mbox{GRACE-B} & X & X & $\bigcirc$ & $\bigcirc$ & $\bigcirc$ \\
        \midrule
        \multirow{5}{*}{Calibration} & \mbox{CryoSat-2} FGM1 & X & X & X & X & $\bigcirc$ \\
        & \mbox{CryoSat-2} FGM2 & X & X & X & X & $\bigcirc$ \\
        & \mbox{CryoSat-2} FGM3 & X & X & X & X & $\bigcirc$ \\
        & \mbox{GRACE-A} & X & X & X & X & $\bigcirc$ \\
        & \mbox{GRACE-B} & X & X & X & X & $\bigcirc$ \\
        \bottomrule
        \multicolumn{7}{l}{\raisebox{-0.5em}{$^{1}$\footnotesize{Entries related to $g_1^0$ B-spline coefficients and platform magnetometer data are zero.}}}
    \end{tabular}
\end{table*}
Nevertheless, we used the full model description in the forward evaluation to compute the residuals.

The iterative procedure described in Eq.~\eqref{eq:iterative_procedure} requires a starting model $\mathbf{m}_0$ to initialize the model parameter estimation. We initialized the internal field model parameters using the corresponding part of \mbox{CHAOS-6-x9}, while we set the external field model parameters to zero. To initialize the Euler angles, we used the values from \mbox{CHAOS-6-x9} in case of \textit{Swarm} and CHAMP satellites, or set the angles to zero in case of \mbox{CryoSat-2} and the GRACE satellite duo. For the calibration parameters, we simply set the offsets and non-orthogonalities to zero and the sensitivities to one over the whole time span. The parameter estimation usually converged after 10\---15 iterations. We also tested other starting models, e.g.~random calibration parameters, but found that our choice had little impact on the converged model parameters other than increasing the number of necessary iterations.

\subsection{Data weighting}
\label{sec:data_weighting}

For the vector components of the non-polar satellite data, we used a covariance matrix that accounts for the attitude uncertainty of the star cameras
\begin{equation}
    \doubleunderline{\mathbf{C}_\mathrm{B23}} = \diag{\sigma^2, \sigma^2 + B^2\psi^2, \sigma^2 + B^2\psi^2}
\end{equation}
with respect to the B23 reference frame defined by unit vectors in the direction of $\mathbf{B}$, $\mathbf{n}\times \mathbf{B}$, and $\mathbf{n}\times(\mathbf{n}\times \mathbf{B})$, where $\mathbf{n}$ is an arbitrary unit vector not parallel to $\mathbf{B}$ that we chose to be the third CRF base vector, $\sigma^2$ is the variance of an isotropic instrument error and $\psi^2$ is the variance associated with random rotations around the three reference axes \cite[]{Holme1996}. Tab.~\ref{tab:satellite_uncertainty} summarizes the values of $\sigma$ and $\psi$ for the different satellite datasets.
\begin{table*}
    \caption{Chosen values of $\sigma$ and $\psi$ for the different satellites. The values under Swarm apply to the data from the two Swarm satellites in this study (Swarm-A and Swarm-B), the values under CryoSat-2 to the data of the three magnetometers (FGM1, FGM2 and FGM3), and the values under GRACE to the data from both GRACE satellites (GRACE-A and GRACE-B).}
    \label{tab:satellite_uncertainty}
    \centering
    \begin{tabular}{ccccc}
    \toprule
     & CHAMP & Swarm & CryoSat-2 & GRACE \\
    \midrule
    $\sigma$ (\si{\nT}) & 2.5 & 2.2 & 6 & 10\\
    $\psi$ (\si{arcsec}) & 10 & 5 & 30 & 100 \\
    \bottomrule
    \end{tabular}
\end{table*}
We scaled the diagonal entries of the covariance matrix with Huber weights \cite[]{Constable1988,Sabaka2004} that we calculated for each component in the B23 reference frame to downweight data points that greatly deviated from the model evaluated at the previous iteration. After inverting and rotating the Huber-weighted covariance matrix of the individual data point into the RTP frame, we arranged them into a block-diagonal matrix completing the desired inverse data covariance matrix $\doubleunderline{\mathbf{C}_\mathrm{d}}^{-1}$. In case of the vector magnitude of the polar satellite data, we simply used $\sigma^2$ scaled with Huber weights as variance. The covariance of the ground observatory SV vector data was derived from detrended residuals to the \mbox{CHAOS-6-x9} model, including the covariance between vector components at a given location.

\subsection{Model regularization}
\label{sec:model_regularization}

The regularization in the form of the matrix $\doubleunderline{\mathbf{\Lambda}}$ in Eq.~\eqref{eq:optimization} is designed to ensure the convergence of the model parameter estimation by limiting the flexibility of the model. The regularization matrix is block diagonal and consists of the blocks $\doubleunderline{\mathbf{\Lambda}_\mathrm{int}}$, $\doubleunderline{\mathbf{\Lambda}_\mathrm{ext}}$, and $\doubleunderline{\mathbf{\Lambda}_\mathrm{cal}}$, which regularized the internal, external, and the calibration parameters, respectively. We did not regularize the Euler angles, such that corresponding blocks in the regularization matrix are zero.

Turning to the internal part of the model, following the example of earlier models in the CHAOS series, we designed a regularization based on the square of the third time-derivative of the radial field component $B_r$ integrated over the core mantle boundary (CMB) and averaged over the entire model time span
\begin{equation}
    \label{eq:internal_3rd_reg}
    \langle\dddot{B}_r^2\rangle = \frac{1}{4\uppi(t_\mathrm{e} - t_\mathrm{s})}\int_{t_\mathrm{s}}^{t_\mathrm{e}}\int_{\Omega(c)}\left(\frac{\partial^3 B_r}{\partial t^3}\right)^2\mathrm{d}\Omega\mathrm{d}t
\end{equation}
where $c=\SI{3485.0}{\kilo\meter}$ is the chosen spherical reference radius of the CMB, $\Omega(c)$ denotes the CMB given as the spherical surface of radius $c$, and $\mathrm{d}\Omega=\sin\theta\mathrm{d}\theta\mathrm{d}\phi$ is the surface element for the integration. Furthermore, we set up a regularization of the internal field based on the square of the second time-derivative of the radial component integrated over the CMB at the model start time $t_\mathrm{s}$
\begin{equation}
    \label{eq:internal_2rd_reg}
    \langle\ddot{B}_r^2(t_\mathrm{s})\rangle = \frac{1}{4\uppi}\int_{\Omega(c)} \left(\left.\frac{\partial^2 B_r}{\partial t^2}\right|_{t=t_\mathrm{s}}\right)^2\mathrm{d}\Omega,
\end{equation}
and similarly for the end time by replacing $t_\mathrm{s}$ with $t_\mathrm{e}$. Returning to Eq.~\eqref{eq:internal_3rd_reg}, thanks to the orthogonality of spherical harmonics on the surface of the sphere, carrying out the spatial integration leads to
\begin{equation}
    \big\langle\dddot{B}_r^2\big\rangle = \sum_{n=1}^{N_\mathrm{int}}\bigg(w_\Omega(n)\sum_{m=0}^n \left(\big\langle \dddot{g}_n^m(t)^2\big\rangle_t + \big\langle \dddot{h}_n^m(t)^2\big\rangle_t\right)\bigg)
\end{equation}
where $w_\Omega=\frac{(n+1)^2}{2n+1}\left(\frac{a}{c}\right)^{2n+4}$ is a spatial factor that follows from the surface integration and $\langle\cdot\rangle_t=\frac{1}{t_\mathrm{e} - t_\mathrm{s}}\int_{t_\mathrm{s}}^{t_\mathrm{e}}\mathrm{d}t$ denotes the time average. Utilizing the fact that the time-dependence of the Gauss coefficients is given by sixth-order B-splines, terms such as
\begin{equation}
    \begin{aligned}
        \big\langle \dddot{g}_n^m(t)^2\big\rangle_t &= \sum_{j,j'} g_{n,j}^m g_{n,j'}^m \big\langle \dddot{B}_{6,j}(t)\dddot{B}_{6,j'}(t)\big\rangle_t\\
        &= \sum_{j,j'} g_{n,j}^m g_{n,j'}^m A_{jj'}\\
        &= (\mathbf{g}_n^m)^\mathrm{T}\doubleunderline{\mathbf{A}_t}\mathbf{g}_n^m
    \end{aligned}
\end{equation}
can be written as a quadratic form in $\mathbf{g}_n^m = (g_{n, 1}^m, g_{n, 2}^m, \dots)^\mathrm{T}$, the vector of the spline coefficients of $g_n^m$, using the matrix $\doubleunderline{\mathbf{A}_t}$ that has entries corresponding to the time averages of products of the third time-derivative of the B-splines. While the time-derivatives of the B-splines are known analytically, we approximated the time average numerically by a Riemann sum of rectangles. A similar computation of Eq.~\eqref{eq:internal_2rd_reg}, now evaluating the derivatives only at the endpoints instead of averaging in time, yields matrices
$\big(\doubleunderline{\mathbf{A}_{t_\mathrm{s}}}\big)_{jj'}=\ddot{B}_{6,j}(t_\mathrm{s})\ddot{B}_{6,j'}(t_\mathrm{s})$ and $\big(\doubleunderline{\mathbf{A}_{t_\mathrm{e}}}\big)_{jj'}=\ddot{B}_{6,j}(t_\mathrm{e})\ddot{B}_{6,j'}(t_\mathrm{e})$. Finally, based on the physical quantities in Eqs.~\eqref{eq:internal_3rd_reg} and \eqref{eq:internal_2rd_reg}, we devised a block-diagonal regularization matrix for the internal magnetic field model
\begin{equation}
    \label{eq:internal_reg_all}
    \begin{aligned}
       \doubleunderline{\mathbf{\Lambda}_\mathrm{int}} = \underset{n,m}{\mathrm{diag}}\bigg(& w_\Omega(n)w_m(m)w_\mathrm{tp}(n)\bigg.\\
       \bigg. &\cdot\big(\lambda_t\doubleunderline{\mathbf{A}_t} + \lambda_{t_\mathrm{s}}\doubleunderline{\mathbf{A}_{t_\mathrm{s}}} + \lambda_{t_\mathrm{e}}\doubleunderline{\mathbf{A}_{t_\mathrm{e}}}\big)\bigg)
    \end{aligned}
\end{equation}
where $n$ and $m$ run over the degree and order in the spherical harmonic expansion of the internal field in Eq.~\eqref{eq:potential_int}; $w_m(m)$ and $w_\mathrm{tp}(n)$ are functions which control the regularization strength based on the degree and order of the internal Gauss coefficients; $\lambda_t$, $\lambda_{t_\mathrm{s}}$, and $\lambda_{t_\mathrm{e}}$ are parameters that, respectively, set the regularization strength over the entire model time span, at the model start time and end time. Following \cite{Finlay2020}, in order to relax the regularization at higher spherical harmonic degree, we defined $w_\mathrm{tp}(n)$ as a tapered window which gradually reduces from one to 0.005
\begin{equation}
    w_\mathrm{tp}(n) =
    \left\{
    \begin{aligned}
        & 1, \quad n < n_\mathrm{min}\\
        & \tau(n), \quad n_\mathrm{min} \leq n \leq n_\mathrm{max} \\
        & 0.005, \quad n > n_\mathrm{max}
    \end{aligned}
    \right.
\end{equation}
where $n_\mathrm{min} = 3$ and $n_\mathrm{max} = 6$ are the chosen limits of a half-cosine taper
\begin{equation}
    \tau(n) = \frac{0.995}{2}\bigg[ 1 + \cos\left(\pi\frac{n - n_\mathrm{min}}{n_\mathrm{max} - n_\mathrm{min}}\right)\bigg] + 0.005.
\end{equation}
In contrast to \cite{Finlay2020}, who used $n_\mathrm{max} = 11$ to achieve stable power spectra with more power in the time-dependence of the high-degree coefficients without causing instabilities, we were able to further decrease the upper limit of the taper. The magnetospheric and ionospheric field and their induced counterparts may also cause the estimation of the internal field parameters to become unstable. Our experience shows that it is typically the zonal harmonics that become unstable first if the regularization is not sufficiently strong. Therefore, in addition to the degree-dependent temporal regularization, there is a special treatment of zonal and non-zonal spherical harmonics based on
\begin{equation}
    w_m(m) =
    \left\{
    \begin{aligned}
        & \lambda_0, \quad m = 0\\
        & \lambda_m, \quad m \neq 0.
    \end{aligned}
    \right.
\end{equation}
Note that the regularization of the internal field model only constrains the time-derivatives of the field but not the field itself.

Turning to the external part of the model, we regularized only the bin-to-bin variability of the three RC baseline corrections $\Delta q_{1, \mathrm{SM}}^{0}$, $\Delta q_{1, \mathrm{SM}}^{1}$, and $\Delta s_{1, \mathrm{SM}}^{1}$ in Eq.~\eqref{eq:potential_sm} using a quadratic form in the first forward difference of neighboring bins. The forward difference was calculated with the matrix
\begin{equation}
    \doubleunderline{\mathbf{D}} =
    \frac{1}{t_\mathrm{e} - t_\mathrm{s}}
    \begin{pmatrix}
    -1 & 1 & & \\
       & \ddots & \ddots & \\
       &  & -1 & 1
    \end{pmatrix},
\end{equation}
whose number of columns is equal to the number of bins that comprise each RC-baseline correction. Taken together, the regularization matrix for all parameters related to the external field model reads
\begin{equation}
    \doubleunderline{\mathbf{\Lambda}_\mathrm{ext}} = \diag{0, \dots, 0,\lambda_\mathrm{ext}\doubleunderline{\mathbf{I}_3}\otimes\doubleunderline{\mathbf{D}_2} ,0 , \dots, 0},
\end{equation}
where $\otimes$ is the Kronecker product, $\doubleunderline{\mathbf{I}_3}$ is the unit matrix of size three corresponding to the three RC-baseline corrections, $\doubleunderline{\mathbf{D}_2} =\doubleunderline{\mathbf{D}}^\mathrm{T}\doubleunderline{\mathbf{D}}$ is the coefficients matrix that determines the quadratic form, additional zeros on the diagonal indicate the other unregularized model parameters of the external field, and $\lambda_\mathrm{ext}$ is the chosen regularization parameter.

Turning to the calibration parameters, we regularized a quadratic form in the bin-to-bin variability of each calibration parameter for the five platform magnetometers (three on \mbox{CryoSat-2} and one on each of the two GRACE satellites). The regularization matrix $\doubleunderline{\mathbf{\Lambda}_\mathrm{cal}}$ is block-diagonal with each block $\doubleunderline{\mathbf{\Lambda}_{\mathrm{cal},i}}$, $i=1,\dots, 5$, corresponding to the calibration parameters for each of the five platform magnetometers. The regularization matrix can be written as
\begin{equation}
    \begin{aligned}
        \doubleunderline{\mathbf{\Lambda}_\mathrm{cal}} &= \diag{\doubleunderline{\mathbf{\Lambda}_{\mathrm{cal},1}},\dots, \doubleunderline{\mathbf{\Lambda}_{\mathrm{cal},5}}} \\
        \doubleunderline{\mathbf{\Lambda}_{\mathrm{cal},i}} &= \diag{\lambda_{b,i}, \lambda_{s,i}, \lambda_{u,i}}\otimes\doubleunderline{\mathbf{I}_3}\otimes\doubleunderline{\mathbf{D}_2},
    \end{aligned}
\end{equation}
where we define the regularization parameters $\lambda_{b,i}$, $\lambda_{s,i}$ and $\lambda_{u,i}$ to control the temporal smoothness of the offsets, sensitivities, and non-orthogonalities, respectively.

\section{Results and discussion}
\label{sec:results}

We built two geomagnetic field models which span 10 years from the 1st of January 2008 to the 31st of December 2018, but differ in the use of platform magnetometer data to constrain the field model parameters.

The first model, \mbox{Model-A}, was derived with data from the \mbox{Swarm-A}, \mbox{Swarm-B}, and CHAMP satellites, and the monthly SV data from ground observatories. It served as a reference model, which allowed us to identify differences to models which were derived using platform magnetometer data in addition. Considering the model parameterization, regularization, and estimation, \mbox{Model-A} is very similar to the CHAOS model series. In fact, the parameterization of the geomagnetic field and the alignment parameters of the satellite data are identical, except for the lower truncation degree of the internal field and the longer bins of the alignment parameters and RC-baseline corrections in \mbox{Model-A}. A notable difference is the use of gradient data in the CHAOS model. The strong temporal regularization of the high-degree Gauss coefficients of the time-dependent internal field has been relaxed in the newly released CHAOS-7 model through a taper, which we also used here. For \mbox{Model-A}, we tuned the regularization, such that the model parameters matched the ones of the \mbox{CHAOS-6-x9} model as close as possible. Tab.~\ref{tab:regularization_parameters} shows the numerical values of the regularization parameters.
\begin{table*}
    \caption{Chosen numerical values of the regularization parameters. The values are valid for all the models built in this paper insofar as the regularization terms are applicable to the specific model.}
    \label{tab:regularization_parameters}
    \begin{tabular}{llc}
        \toprule
        & & Regularization parameter\\
        \multicolumn{2}{l}{Description of the model parameters} & \\
        \midrule
        \multirow{2}{*}{Internal field} & \multirow{2}{*}{Time-dependent} & $\lambda_t = \SI{1.0}{\left(\frac{\nT}{\year\cubed}\right)^{-2}}$, $\lambda_{t_\mathrm{s}} = \SI{0.03}{\left(\frac{\nT}{\year\squared}\right)^{-2}}$, $\lambda_{t_\mathrm{e}} = \SI{0.03}{\left(\frac{\nT}{\year\squared}\right)^{-2}}$,\\
        & & $\lambda_\mathrm{0} = 60$, $\lambda_\mathrm{m} = 0.65$\\
        \midrule
        External field & RC-baseline corrections & $\lambda_\mathrm{ext} = \SI{4e5}{\big(\frac{\nT}{\year}\big)^{-2}}$ \\
        \midrule
        \multirow{5}{*}{Calibration$^{1}$} & \mbox{CryoSat-2} FGM1 & $\lambda_b = \SI{9.1e2}{\big(\frac{\eu}{\year}\big)^{-2}}$, $\lambda_s = \SI{9.1e10}{\big(\frac{\eu}{\nT\year}\big)^{-2}}$, $\lambda_u = \SI{2.8e2}{\big(\frac{1\degree}{\year}\big)^{-2}}$ \\
        & \mbox{CryoSat-2} FGM2 & $\lambda_b = \SI{9.1e2}{\big(\frac{\eu}{\year}\big)^{-2}}$, $\lambda_s = \SI{9.1e10}{\big(\frac{\eu}{\nT\year}\big)^{-2}}$, $\lambda_u = \SI{2.8e2}{\big(\frac{1\degree}{\year}\big)^{-2}}$ \\
        & \mbox{CryoSat-2} FGM3 & $\lambda_b = \SI{9.1e2}{\big(\frac{\eu}{\year}\big)^{-2}}$, $\lambda_s = \SI{9.1e10}{\big(\frac{\eu}{\nT\year}\big)^{-2}}$, $\lambda_u = \SI{2.8e2}{\big(\frac{1\degree}{\year}\big)^{-2}}$ \\
        & \mbox{GRACE-A} & $\lambda_b = \SI{1.2e3}{\big(\frac{\eu}{\year}\big)^{-2}}$, $\lambda_s = \SI{1.2e13}{\big(\frac{\eu}{\nT\year}\big)^{-2}}$, $\lambda_u = \SI{3.7e8}{\big(\frac{1\degree}{\year}\big)^{-2}}$ \\
        & \mbox{GRACE-B} & $\lambda_b = \SI{1.2e3}{\big(\frac{\eu}{\year}\big)^{-2}}$, $\lambda_s = \SI{1.2e13}{\big(\frac{\eu}{\nT\year}\big)^{-2}}$, $\lambda_u = \SI{3.6e8}{\big(\frac{1\degree}{\year}\big)^{-2}}$\\
        \bottomrule
        \multicolumn{3}{l}{\raisebox{-0.5em}{$^{1}$\footnotesize{Not applicable to \mbox{Model-A}, which was not derived from platform magnetometer data.}}}
    \end{tabular}
\end{table*}

The second model, \mbox{Model-B}, is our preferred model and was derived with data from \mbox{Swarm-A}, \mbox{Swarm-B}, CHAMP, monthly ground observatory SV data, and, as opposed to \mbox{Model-A}, platform magnetometer data from \mbox{CryoSat-2} FGM1, \mbox{CryoSat-2} FGM2, \mbox{CryoSat-2} FGM3, \mbox{GRACE-A}, and \mbox{GRACE-B}. In addition to \mbox{Model-A} and \mbox{Model-B}, we built test models in a series of experiments to investigate the effect of platform magnetometer data on the estimation of the geomagnetic field model. Details of the test models are given below. The regularization parameters are the same for all the presented models, i.e., \mbox{Model-A}, \mbox{Model-B}, and the test models.

\subsection{Fit to satellite data and ground observatory SV data}

\begin{figure*}
    \includegraphics[width=\figurewidth]{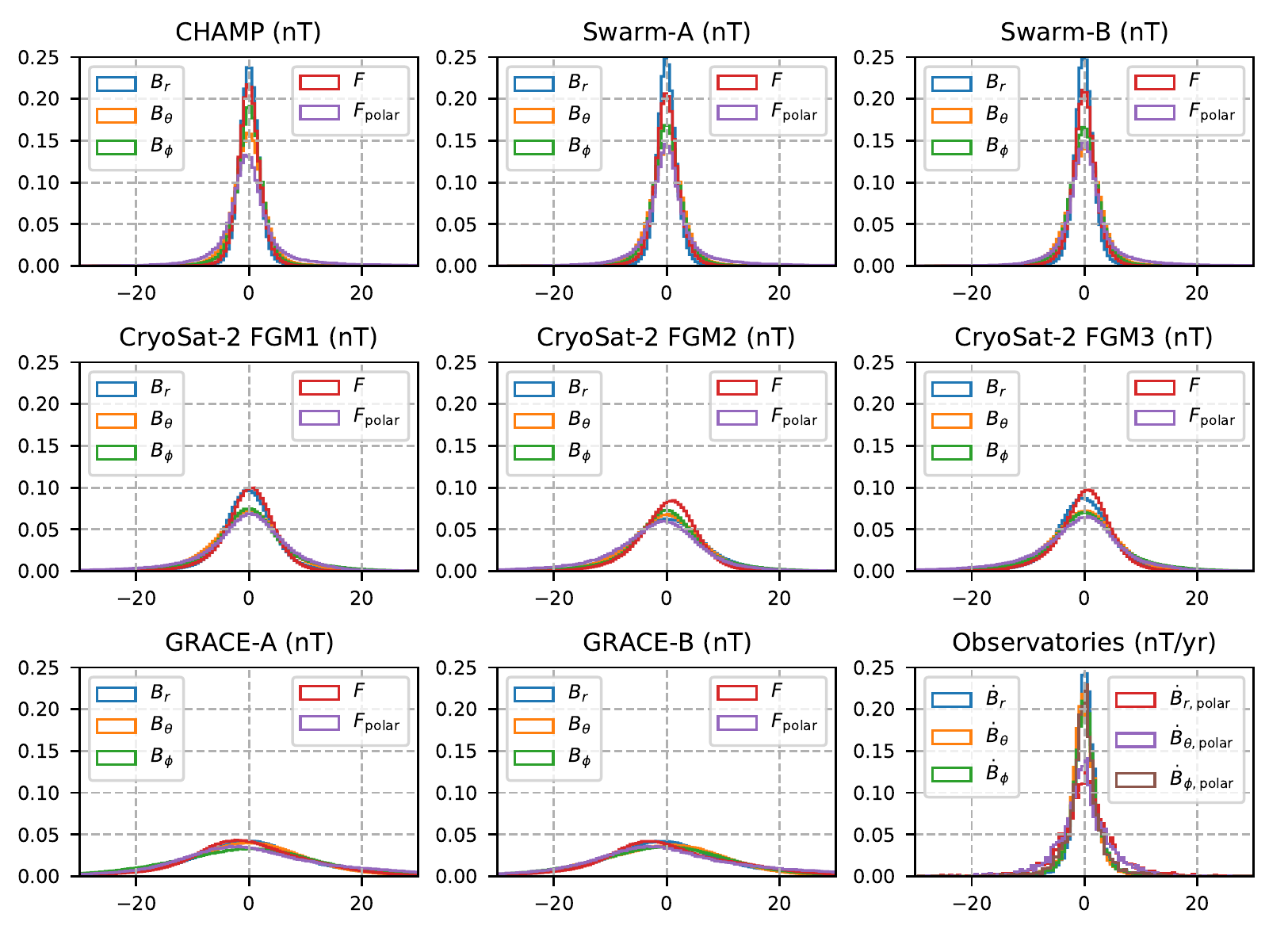}
    \caption{Histograms of the residuals of each satellite and ground observatory SV data using \mbox{Model-B}. The histograms have been normalized to have unit area. Computed statistics are shown in Tab.~\ref{tab:statistics_satellites} for the satellite data and Tab.~\ref{tab:statistics_ground} for the ground observatory SV data.}
    \label{fig:histograms_satellites}
\end{figure*}

We begin with reporting on the fit of \mbox{Model-B} to the satellite data and ground observatory SV data. The histograms of the scalar and vector residuals for each dataset are shown in Fig.~\ref{fig:histograms_satellites}.
The residuals of \mbox{Swarm-A}, \mbox{Swarm-B}, CHAMP and the ground observatories show narrow and near-zero centered peaks, which demonstrate the high-quality and low-noise level of these datasets. In contrast, the peaks are broader for \mbox{CryoSat-2} and even more in the case of GRACE, which is, as expected, due to the higher data noise level. By separating the residuals poleward of \ang{\pm 55} QD latitude from the ones equatorward, we find that peaks are broader at polar QD latitudes for all datasets, which is a result of unmodeled magnetic signal of the polar ionospheric current system. Also, the histograms of the GRACE residuals are biased toward negative values. Upon further investigation, we found a local time-dependence especially visible in the scalar residuals, which could indicate that signals from solar array and battery currents have not been fully removed from the GRACE datasets used here. The residual statistics are summarized in Tab.~\ref{tab:statistics_satellites} for the satellite data and Tab.~\ref{tab:statistics_ground} for the ground observatory SV data.
\begin{table*}
    \caption{Number $N$, Huber-weighted mean, and standard deviation $\sigma$ computed from the residuals of the satellite data for each vector component and split into polar (poleward \ang{\pm 55}) and non-polar (equatorward \ang{\pm 55}) QD latitudes. Note that non-polar scalar data were not used in the model parameter estimation\----statistics are only shown for completeness.}
    \label{tab:statistics_satellites}
    \begin{tabular}{lllrrr}
    \toprule
            &       &     &        N &  mean (\si{nT}) &  $\sigma$ (\si{nT}) \\
    Dataset & Quasi-dipole latitude & Component &          &                 &                     \\
    \midrule
    \multirow{5}{*}{CHAMP} & \multirow{4}{*}{Non-polar} & $B_r$ &   707131 &            0.02 &                1.93 \\
            &       & $B_\theta$ &   707131 &           -0.11 &                2.84 \\
            &       & $B_\phi$ &   707131 &            0.03 &                2.32 \\
            &       & $F$ &   707131 &            0.01 &                1.93 \\
    \cmidrule{2-6}
            & Polar & $F$ &   200084 &           -0.02 &                5.10 \\
    \cmidrule{1-6}
    \multirow{5}{*}{CryoSat-2 FGM1} & \multirow{4}{*}{Non-polar} & $B_r$ &   958362 &           -0.06 &                4.39 \\
            &       & $B_\theta$ &   958362 &           -0.31 &                5.76 \\
            &       & $B_\phi$ &   958362 &            0.06 &                6.49 \\
            &       & $F$ &   958362 &            0.06 &                4.18 \\
    \cmidrule{2-6}
            & Polar & $F$ &   331097 &           -0.28 &                7.56 \\
    \cmidrule{1-6}
    \multirow{5}{*}{CryoSat-2 FGM2} & \multirow{4}{*}{Non-polar} & $B_r$ &   958362 &           -0.03 &                6.42 \\
            &       & $B_\theta$ &   958362 &           -0.29 &                6.01 \\
            &       & $B_\phi$ &   958362 &            0.07 &                6.55 \\
            &       & $F$ &   958362 &            0.18 &                4.86 \\
    \cmidrule{2-6}
            & Polar & $F$ &   331097 &           -1.70 &                8.21 \\
    \cmidrule{1-6}
    \multirow{5}{*}{CryoSat-2 FGM3} & \multirow{4}{*}{Non-polar} & $B_r$ &   958362 &           -0.07 &                4.76 \\
            &       & $B_\theta$ &   958362 &           -0.23 &                5.71 \\
            &       & $B_\phi$ &   958362 &            0.04 &                6.80 \\
            &       & $F$ &   958362 &            0.12 &                4.35 \\
    \cmidrule{2-6}
            & Polar & $F$ &   331097 &           -1.01 &                7.86 \\
    \cmidrule{1-6}
    \multirow{5}{*}{GRACE-A} & \multirow{4}{*}{Non-polar} & $B_r$ &  1082071 &           -0.12 &               11.40 \\
            &       & $B_\theta$ &  1082071 &           -0.24 &               10.48 \\
            &       & $B_\phi$ &  1082071 &           -0.79 &               13.57 \\
            &       & $F$ &  1082071 &           -0.16 &               10.59 \\
    \cmidrule{2-6}
            & Polar & $F$ &   356988 &            0.32 &               15.56 \\
    \cmidrule{1-6}
    \multirow{5}{*}{GRACE-B} & \multirow{4}{*}{Non-polar} & $B_r$ &   997802 &           -0.30 &               11.77 \\
            &       & $B_\theta$ &   997802 &           -0.69 &               11.09 \\
            &       & $B_\phi$ &   997802 &           -0.68 &               12.35 \\
            &       & $F$ &   997802 &            0.02 &               11.53 \\
    \cmidrule{2-6}
            & Polar & $F$ &   331516 &           -0.24 &               15.56 \\
    \cmidrule{1-6}
    \multirow{5}{*}{Swarm-A} & \multirow{4}{*}{Non-polar} & $B_r$ &   817400 &           -0.03 &                1.65 \\
            &       & $B_\theta$ &   817400 &           -0.06 &                2.97 \\
            &       & $B_\phi$ &   817400 &           -0.02 &                2.59 \\
            &       & $F$ &   817400 &           -0.03 &                2.06 \\
    \cmidrule{2-6}
            & Polar & $F$ &   218776 &            0.22 &                4.66 \\
    \cmidrule{1-6}
    \multirow{5}{*}{Swarm-B} & \multirow{4}{*}{Non-polar} & $B_r$ &   809720 &           -0.09 &                1.63 \\
            &       & $B_\theta$ &   809720 &           -0.05 &                3.02 \\
            &       & $B_\phi$ &   809720 &           -0.04 &                2.61 \\
            &       & $F$ &   809720 &           -0.01 &                2.03 \\
    \cmidrule{2-6}
            & Polar & $F$ &   218106 &            0.30 &                4.29 \\
    \bottomrule
    \end{tabular}
\end{table*}
\begin{table*}
    \caption{Number $N$, Huber-weighted mean, and standard deviation $\sigma$ computed from the residuals of the monthly ground observatory SV data for each component and split into polar (poleward \ang{\pm 55}) and non-polar (equatorward \ang{\pm 55}) QD latitudes.}
    \label{tab:statistics_ground}
    \begin{tabular}{lllrrr}
    \toprule
                 &       &                &      N &  mean (\si{\nT/\year}) &  $\sigma$ (\si{\nT/\year}) \\
    Dataset & Quasi-dipole latitude & Component &        &                        &                            \\
    \midrule
    \multirow{6}{*}{Observatories} & \multirow{3}{*}{Non-polar} & $\dot{B}_r$ &  11348 &                   0.20 &                       2.09 \\
                 &       & $\dot{B}_\theta$ &  11348 &                  -0.18 &                       2.26 \\
                 &       & $\dot{B}_\phi$ &  11348 &                   0.06 &                       2.43 \\
    \cmidrule{2-6}
                 & \multirow{3}{*}{Polar} & $\dot{B}_r$ &   3609 &                   0.22 &                       4.43 \\
                 &       & $\dot{B}_\theta$ &   3609 &                  -0.19 &                       4.21 \\
                 &       & $\dot{B}_\phi$ &   3609 &                  -0.08 &                       2.85 \\
    \bottomrule
    \end{tabular}
\end{table*}

\begin{figure*}
    \includegraphics[width=\figurewidth]{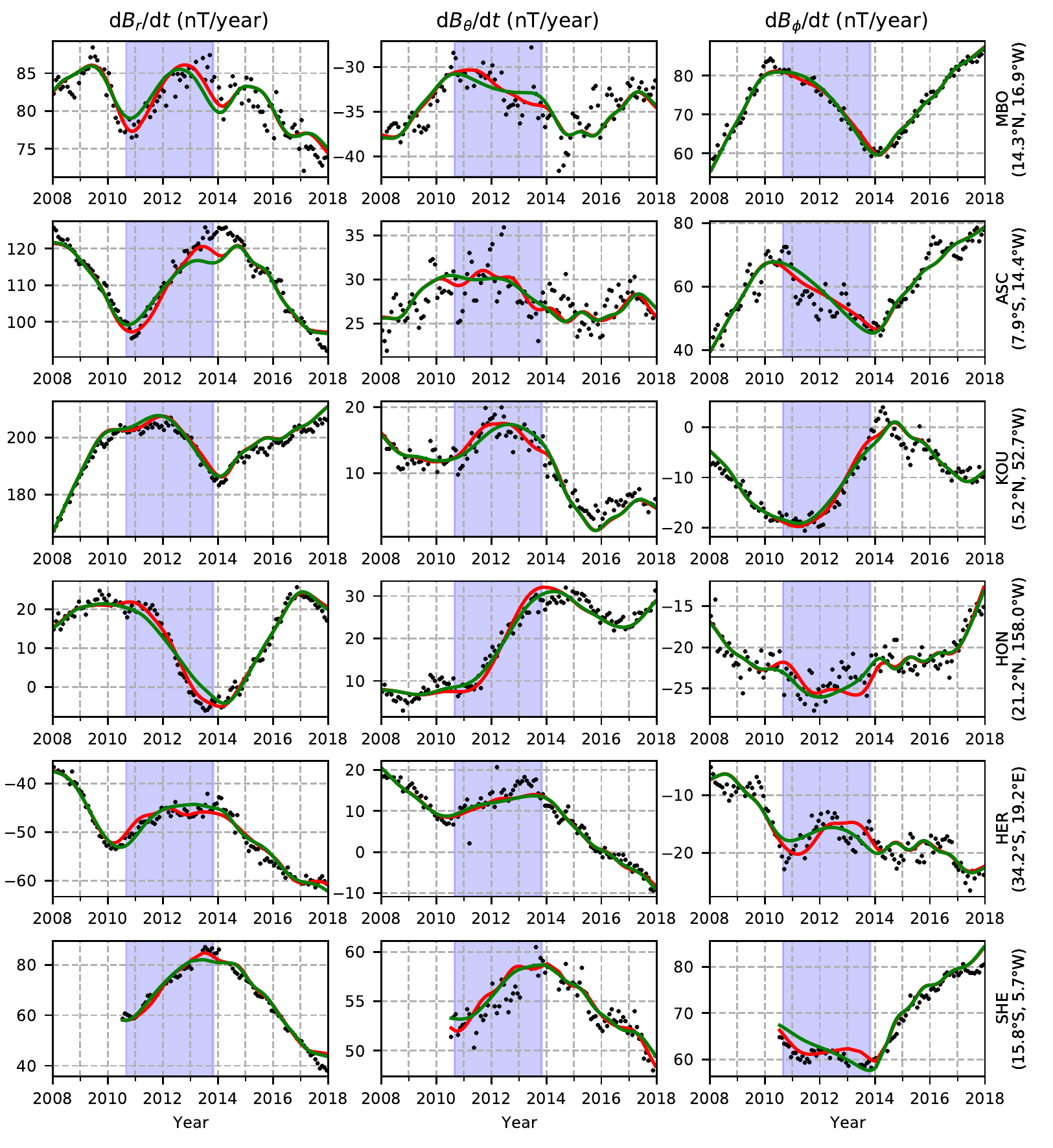}
    \caption{Examples of time-series of monthly ground observatory SV data (black dots) and modeled SV using \mbox{Model-A} (green lines) and \mbox{Model-B} (red lines). The observatory names are MBour (MBO), Ascension (ASC), Kourou (KOU), Honolulu (HON), Hermanus (HER), and Saint Helena (SHE). The SV data of SHE are an independent dataset not used in the inversion. The gap period between CHAMP and \textit{Swarm} is indicated as a blue shaded region (Sep 2010 to Nov 2013).}
    \label{fig:timeseries_ground}
\end{figure*}

Fig.~\ref{fig:timeseries_ground} shows the time-series of the SV components at six chosen ground observatories together with the computed values from \mbox{Model-A} and \mbox{Model-B}.
Overall, the fit of \mbox{Model-A} and \mbox{Model-B} to the ground observatory SV data is good, as expected, for the first five observatory SV shown since these data were used in the model parameter estimation. The computed values of \mbox{Model-A} and \mbox{Model-B} differ especially during the gap from 2010 to 2014, where \mbox{Model-B} can make use of platform magnetometer data in addition to the ground observatory SV data, while \mbox{Model-A} only relies on the ground observatories. That shows that platform magnetometer data contribute to the internal field model especially when there is a lack of calibrated satellite data from CHAMP and \textit{Swarm}. Perhaps even more convincing is the performance of both models when compared to a dataset not used in the inversion. With the SV data from Saint Helena, we show such an independent dataset in the last row of Fig.~\ref{fig:timeseries_ground}. Although both models fit Saint Helena well, \mbox{Model-B} performs slightly better in the radial SV in 2013 and the azimuthal SV at least in the first half of the gap period, until 2012.

To summarize, with \mbox{Model-B} we built a model that fits both the satellite and ground observatory SV data to a satisfactory level, which shows that platform magnetometer data can be successfully used in geomagnetic field modeling.

\subsection{Calibration parameters}

\begin{figure*}
    \includegraphics[width=\figurewidth]{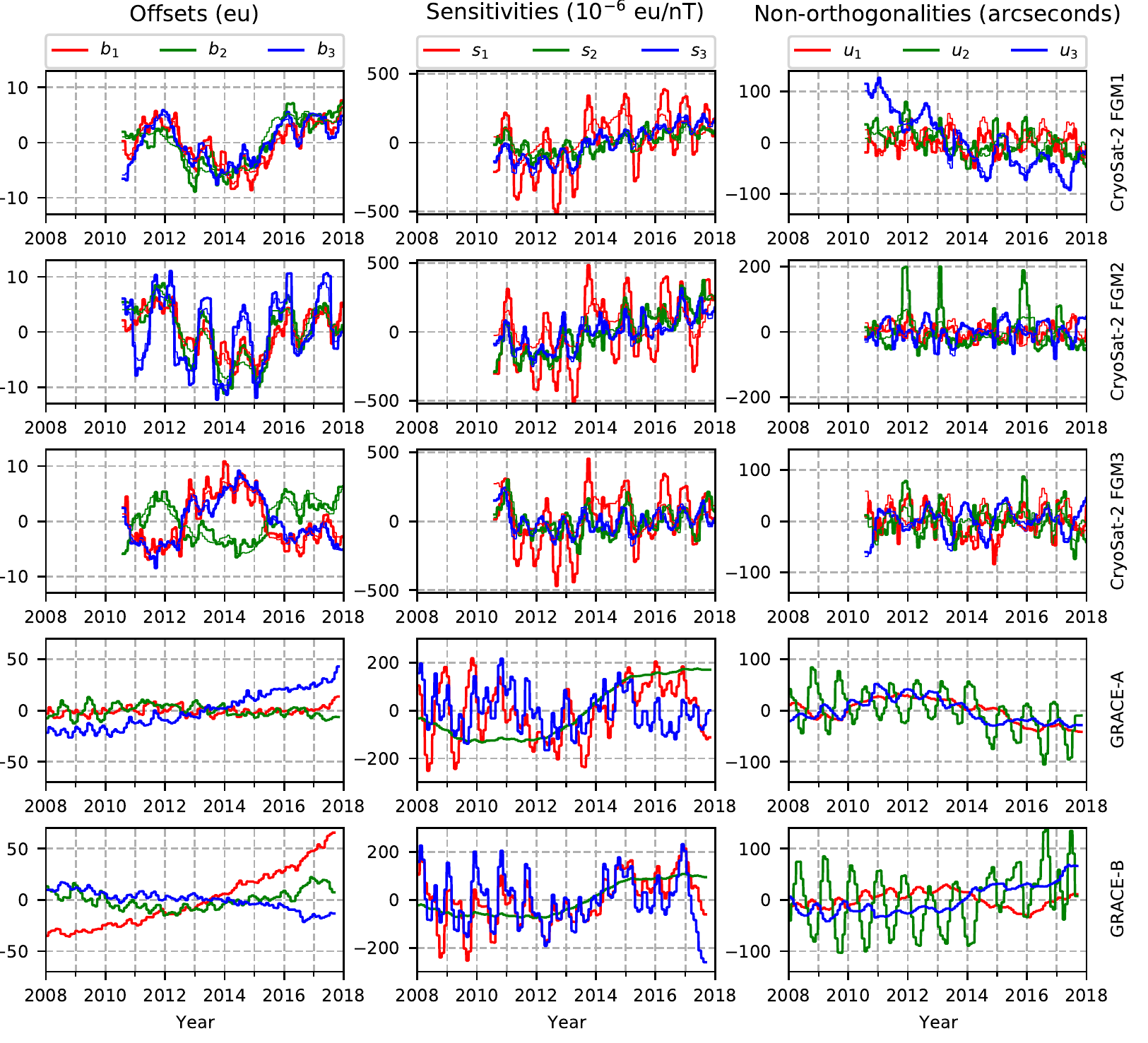}
    \caption{Time series of the calibration parameters of \mbox{Model-B} for each platform magnetometer dataset (thick lines) and calibration parameters of \cite{Olsen2020} for CryoSat-2 (thin lines). The respective mean values in time were removed and are shown in Tab.~\ref{tab:calibration_means}.}
    \label{fig:calibration_parameters}
\end{figure*}

We document the estimated calibration parameters of each platform magnetometer dataset by showing the time-series in Fig.~\ref{fig:calibration_parameters} and the respective mean values in Tab.~\ref{tab:calibration_means}.
\begin{table*}
    \caption{Mean values of the calibration parameters for each platform magnetometer dataset. The time-series are shown in Fig.~\ref{fig:calibration_parameters}.}
    \label{tab:calibration_means}
    \begin{tabular}{lccccccccc}
        \toprule
        {} &  $b_1$ &   $b_2$ &   $b_3$ &    $s_1$ &    $s_2$ &    $s_3$ &          $u_1$ &          $u_2$ &          $u_3$ \\
        {} &   (eu) &    (eu) &    (eu) &  (eu/nT) &  (eu/nT) &  (eu/nT) & (\si{\degree}) & (\si{\degree}) & (\si{\degree}) \\
        Dataset        &        &         &         &          &          &          &                &                &                \\
        \midrule
        CryoSat-2 FGM1 &    5.0 &   165.6 &   -10.7 & 1.005178 & 1.004851 & 1.004479 &          0.453 &          0.191 &         -0.336 \\
        CryoSat-2 FGM2 &   77.6 &   -16.6 &    61.8 & 1.004697 & 1.003993 & 1.003427 &         -0.288 &          0.050 &          0.502 \\
        CryoSat-2 FGM3 & -115.2 &   -29.4 &   -44.6 & 1.000863 & 1.005424 & 1.002168 &          0.745 &         -0.045 &         -0.000 \\
        GRACE-A        &  746.4 & -2632.1 & -2310.0 & 1.034238 & 1.032041 & 1.018168 &         -0.251 &         -0.161 &          0.048 \\
        GRACE-B        &  406.0 & -2622.0 & -2005.6 & 1.029785 & 1.026781 & 1.017845 &         -0.056 &         -0.209 &          0.106 \\
        \bottomrule
    \end{tabular}
\end{table*}
In Fig.~\ref{fig:calibration_parameters}, the rows of panels correspond to the \mbox{CryoSat-2} (top three) and GRACE (bottom two) platform magnetometer datasets, and the columns of panels show the offsets (left), sensitivities (middle), and non-orthogonality angles (right). Since \cite{Alken2020} also used magnetic data from the three platform magnetometers on-board \mbox{CryoSat-2}, it is possible to compare the estimated calibration parameters. First, comparing the time-averaged values of the calibration parameters (Tab.~\ref{tab:calibration_means} here and Tab.~4 in \cite{Alken2020}), we find that the non-orthogonalities are equal to within \ang{0.01} and the offsets to within \SI{1}{\eu}. The averaged values of sensitivities are equal to within \SI{1e-4}{\eu/\nT} (notice that \cite{Alken2020} use the reciprocal of the sensitivity). In terms of the temporal variability, we find that our estimated calibration parameters have amplitudes that are smaller, or equal in case of the offsets, which is likely due to a difference in the regularization strength. In Fig.~\ref{fig:calibration_parameters}, we also show the CryoSat-2 calibration parameters of \cite{Olsen2020} for comparison. Again, the calibration parameters are very similar and differ only in the time variations (e.g., $s_1$) due to the choice of the regularization parameters of this study and \cite{Olsen2020}. Given the acceptable fit to the platform magnetometer data and the reasonable temporal variability of the calibration parameters, we conclude that the calibration of the \mbox{CryoSat-2} and GRACE platform magnetometers was successful.

\subsection{Results of the experiments}

We conducted a series of experiments in which we changed the model estimation, parameterization, and data selection with the goal to investigate and document difficulties when dealing with platform magnetometer data in a co-estimation scheme. This section also justifies the modeling strategies that went into the construction of our preferred geomagnetic field model, \mbox{Model-B}.

\begin{figure*}
    \includegraphics[width=\figurewidth]{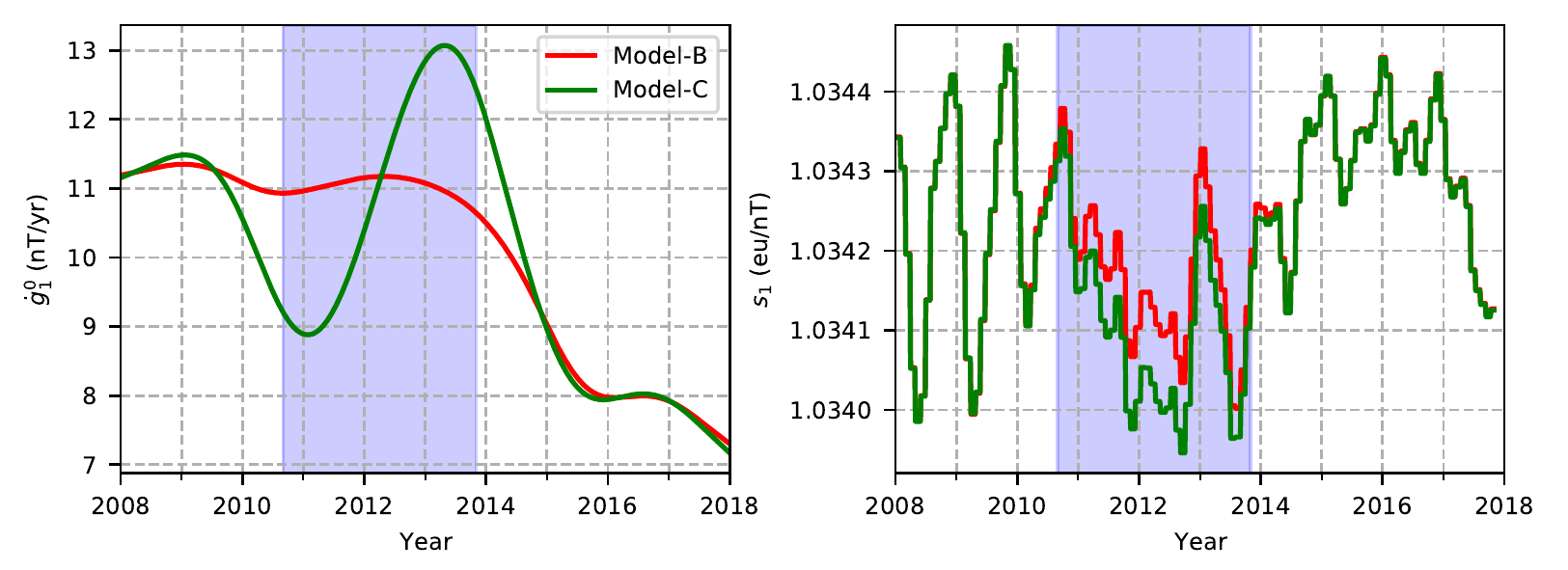}
    \caption{Time-derivative of $g_1^0$ (left) and sensitivity $s_1$ for \mbox{GRACE-A} as computed from \mbox{Model-B} and \mbox{Model-C} (right). For \mbox{Model-C}, we allowed nightside platform magnetometer data to contribute to the estimation of the internal $g_1^0$ Gauss coefficient. The gap period between CHAMP and \textit{Swarm} is indicated as a blue-shaded region (Sep 2010 to Nov 2013).}
    \label{fig:test_g10}
\end{figure*}

In a first experiment, we allowed the nightside platform magnetometer data to participate in the estimation of the axial dipole coefficient of the time-dependent internal field. That is, we derived a test model, \mbox{Model-C}, identical to \mbox{Model-B} but left the matrix of partial derivatives $\doubleunderline{\mathbf{G}}$ unchanged so that the entries corresponding to the B-spline coefficients $g_{1,j}^0$ were non-zero and, thus, the satellite data contributed to the estimation of the internal dipole coefficients.  On the left of Fig.~\ref{fig:test_g10}, we show the time-derivative of $g_1^0$ as a function of time computed with \mbox{Model-B} and \mbox{Model-C}, while, on the right, we show $s_1$ of \mbox{GRACE-A} as an example of the calibration parameters.
In contrast to \mbox{Model-B}, \mbox{Model-C} features a conspicuous detour of the time-derivative of the $g_1^0$ coefficient in the gap between CHAMP and \textit{Swarm} data (blue-shaded region). Although we only show $s_1$ of \mbox{GRACE-A} in Fig.~\ref{fig:test_g10}, we find that all three sensitivities of each platform magnetometer differ in the gap period between \mbox{Model-C} and \mbox{Model-B}. The other internal Gauss coefficients also deviate but to a lesser extent. Interestingly, other model parameters such as the offsets, non-orthogonality angles, Euler angles and external field parameters seem qualitatively unaffected. The same correlation between the internal axial dipole coefficient and the sensitivities has been reported by \cite{Alken2020} who show that this effect can be mitigated either by including large amounts of previously calibrated data or through the use of a regularization that favors a linear time-dependence of the internal dipole during the gap period. Due to the lack of additional calibrated data and our interest in the high-degree SA during the gap that such a regularization affects by redistributing power to higher degrees, we chose to set the dependence of $g_1^0$, the most affected internal Gauss coefficient, on the satellite platform magnetometer data to zero. In other words, we completely relied on the ground observatory SV data and the temporal regularization to estimate the time-dependence of $g_1^0$ in the gap period.

\begin{figure*}
    \includegraphics[width=\figurewidth]{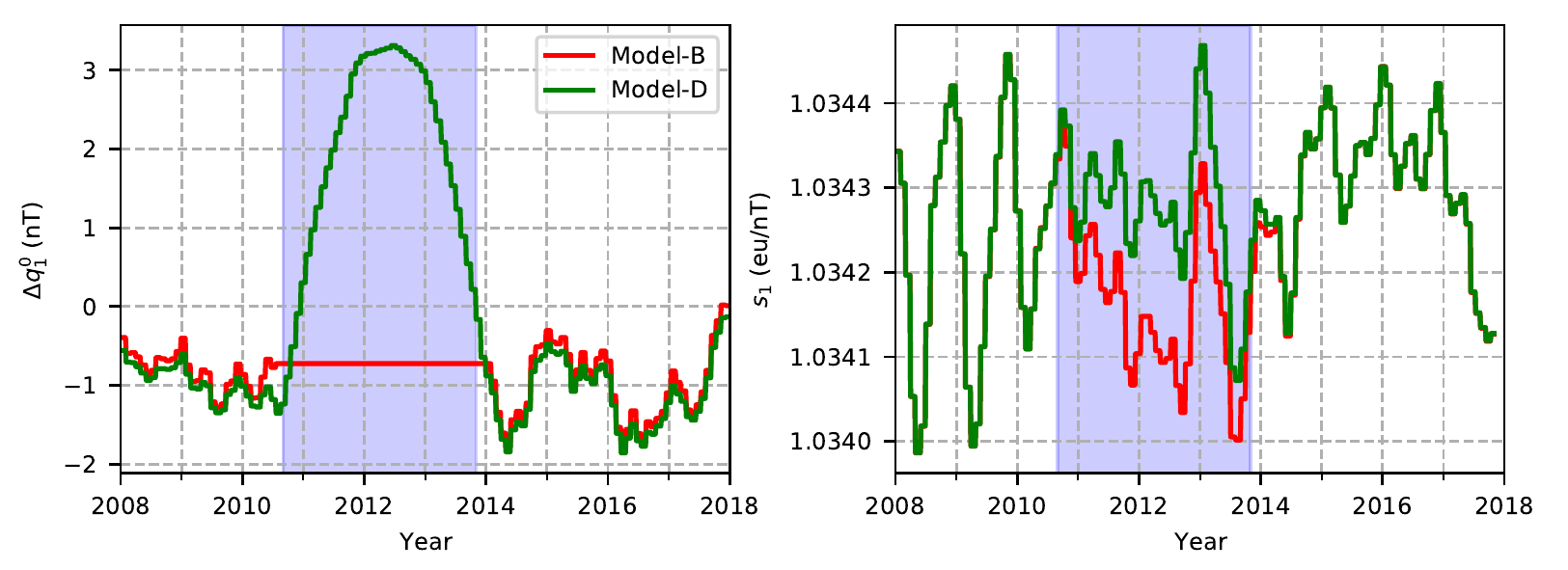}
    \caption{Time series of the RC-baseline correction $\Delta q_1^0$ (left) and sensitivity $s_1$ for \mbox{GRACE-A} as computed from \mbox{Model-B} and \mbox{Model-D} (right). The gap period between CHAMP and \textit{Swarm} is indicated as a blue shaded region (Sep 2010 to Nov 2013). For \mbox{Model-D}, the bins of the RC-baseline corrections are 30 days over the entire model time span, while they were merged to a single bin in the gap period for \mbox{Model-B}.}
    \label{fig:test_SM}
\end{figure*}

In a second experiment, we built a test model, \mbox{Model-D}, which uses 30 day bins of the RC-baseline corrections consistently over the whole model time span in contrast to \mbox{Model-A} and \mbox{Model-B}, which use a single bin spanning the entire gap period. As an example, Fig.~\ref{fig:test_SM} shows the RC-baseline correction $\Delta q_1^0$ on the left and the calibration parameter $s_1$ of \mbox{GRACE-A} on the right, computed with \mbox{Model-D} and \mbox{Model-B}.
In \mbox{Model-D}, $\Delta q_1^0$ has a noticeable peak during the gap period that is much larger in value than the variation during CHAMP or \textit{Swarm} times while the sensitivity is slightly offset to higher values. We find the same behavior for all RC-baseline corrections and calibration parameters, although most prominently for the sensitivities. Again, other model parameters seem unchanged, which indicates that there is a significant correlation between the RC-baseline corrections and the calibration parameters of the platform magnetometers. Using a single bin for the RC-baseline corrections in the gap period helps to reduce that effect. As a final comment regarding \mbox{Model-C} and \mbox{Model-D}, we performed a simulation combining both experiments; that is, we determined $g_1^0$ with the platform magnetomter data and estimated the RC-baseline corrections in 30 day bin over the entire model time span. In this case, we observed deviations from \mbox{Model-B} which were identical to those shown in Figs.~\ref{fig:test_g10} and \ref{fig:test_SM} but, now, affected the internal axial dipole, the RC-baseline corrections, and the sensitivities all at the same time.

In an effort to analyze the relationship between the calibration and the other model parameters in a quantitative manner, we also investigated the model correlations $\rho_{ij} = C_{ij}/\sqrt{C_{ii}C_{jj}}$ based on the entries of the model covariance matrix
\begin{equation}
    \doubleunderline{\mathbf{C}} = \big(\doubleunderline{\mathbf{G}}^\mathrm{T}\doubleunderline{\mathbf{C}_\mathrm{d}}^{-1}\doubleunderline{\mathbf{G}}+\doubleunderline{\mathbf{\Lambda}}\big)^{-1}
\end{equation}
evaluated with the converged model parameters \cite[p.~71]{Tarantola2005}. Unfortunately, the analysis revealed a large number of small correlations, which are difficult to interpret. Therefore, we did not make significant use of it in the modeling and preferred to rely on experiments to guide our modeling strategy.

\begin{figure*}
    \includegraphics[width=\figurewidth]{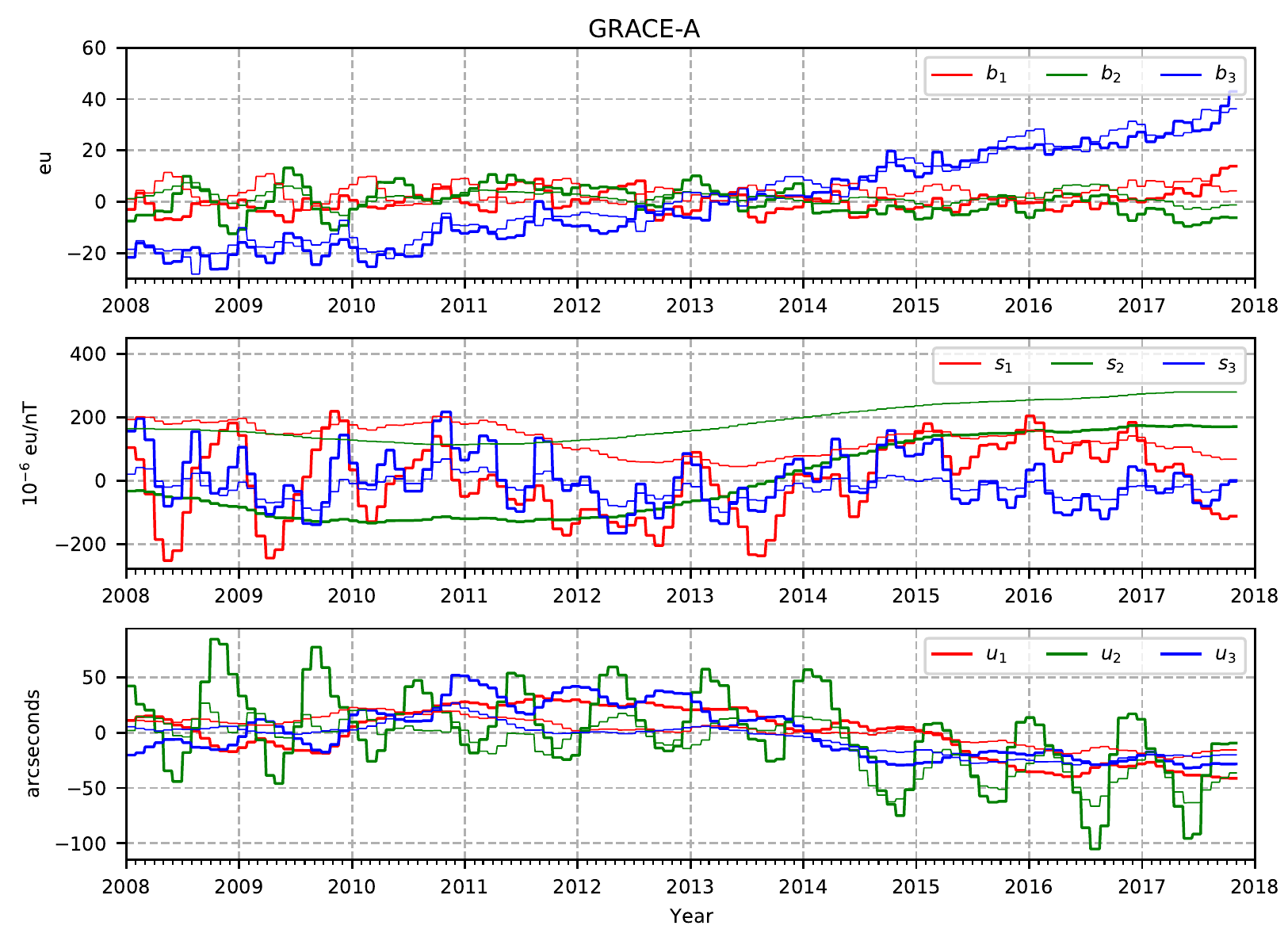}
    \caption{Calibration parameters of \mbox{GRACE-A} computed with \mbox{Model-B} (thick lines) and \mbox{Model-E} (thin lines). We removed the mean values from the calibration parameters as given in Tab.~\ref{tab:calibration_means}.}
    \label{fig:test_dayside}
\end{figure*}

In a final experiment, we derived a test model, \mbox{Model-E}, by only using nightside platform magnetometer data as opposed to \mbox{Model-B}, where the calibration parameters were determined from dayside and nightside platform magnetometer data. Fig.~\ref{fig:test_dayside} shows the calibration parameters for \mbox{GRACE-A} computed with \mbox{Model-B} (thick lines) and \mbox{Model-E} (thin lines).
In the case of \mbox{GRACE-A}, using dayside data to determine the calibration parameters considerably changes the sensitivities and non-orthogonalities as can be seen, for example, when looking at $s_1$, $s_2$ or $u_3$. In particular for $s_2$, there is a vertical shift of approximately \SI{200e-6}{\eu/\nT}, which translates to \SI{10}{\nT} in a magnetic field of \SI{50000}{\nT}. Irrespective of the platform magnetometer, the experiment shows that the local time coverage of the data plays an important role in determining the calibration parameters. The importance of using both day and nightside data becomes clear when appreciating that the orbital plane of the satellites is slowly drifting in local time. Under a possible nightside data selection criteria, the drift leads to the selection of data from either the ascending or descending parts of the orbit at a time. For example, if the ascending node of the orbit is on the nightside, then the platform magnetometer collects data of the magnetic field that mostly points along the direction of flight, in agreement with the predominant dipolar field configuration, until the ascending node crosses over to the dayside placing the descending part of the orbit on the nightside. Now, the observed magnetic field mostly points against the direction of flight. In the case of \mbox{CryoSat-2}, it takes the ascending node 8 months and GRACE around 11 months to traverse the nightside, which is longer than the monthly bins used for estimating the calibration parameters. Hence, the data of each bin will be collected either from the ascending or descending nodes with the respective bias of the field direction. Instead, by using both nightside and dayside, we ensured that the data within each bin covered a broad range of local times to excite the platform magnetometer from various directions, which we believe improves the estimation of the calibration parameters. Nevertheless, we did not use any dayside data to constrain the geomagnetic field model since we do not account for the strong ionospheric sources on the dayside. Those ionospheric sources, however, may contaminate the calibration parameters.

\subsection{Secular acceleration}

One motivation for using platform magnetometer data has been the growing interest in SA pulses, enhancements of the SA that occur on sub-decadal time scales and are seen most prominently at low latitudes. These pulses have been reported by several studies \cite[]{Olsen2007,Chulliat2010, Chulliat2014} and are thought to reflect the dynamical processes in the Earth's outer core. To further study SA pulses and the SA in general, accurate internal field models are needed, which rely on long and continuous time-series of satellite data to give a global picture. When supplemented with high quality satellite data, platform magnetometer data may play an important role in providing those models.

\begin{figure*}
    \includegraphics[width=\figurewidth]{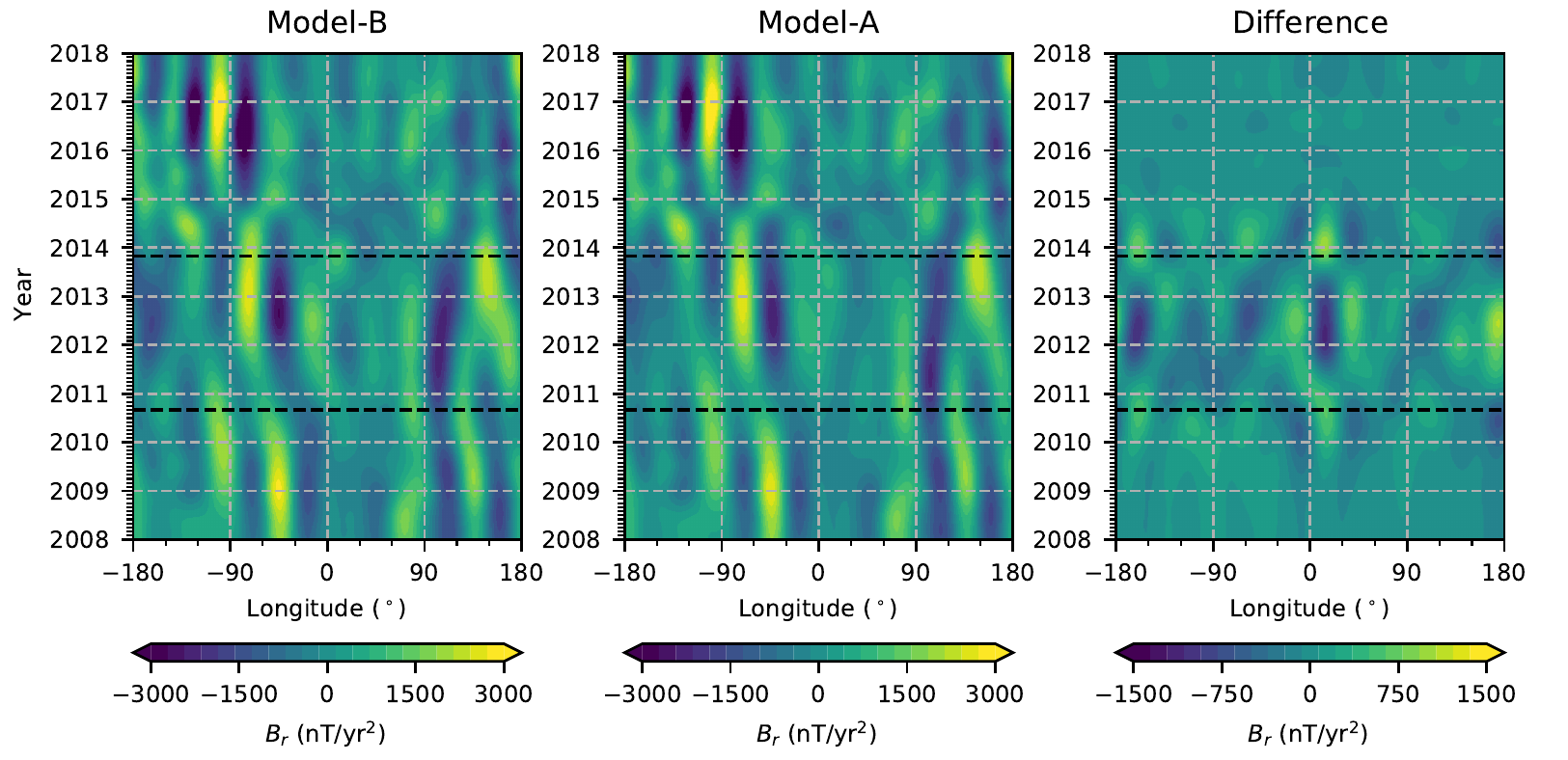}
    \caption{Time-longitude maps of the radial SA up to degree 10 on the Equator at the CMB as computed with \mbox{Model-B} (left), \mbox{Model-A} (center) and their difference, \mbox{Model-B} minus \mbox{Model-A} (right). The gap period between CHAMP and \textit{Swarm} is in between the black dashed lines (Sep 2010\---Nov 2013).}
    \label{fig:SA_maps}
\end{figure*}

\begin{figure*}
    \includegraphics[width=\figurewidth]{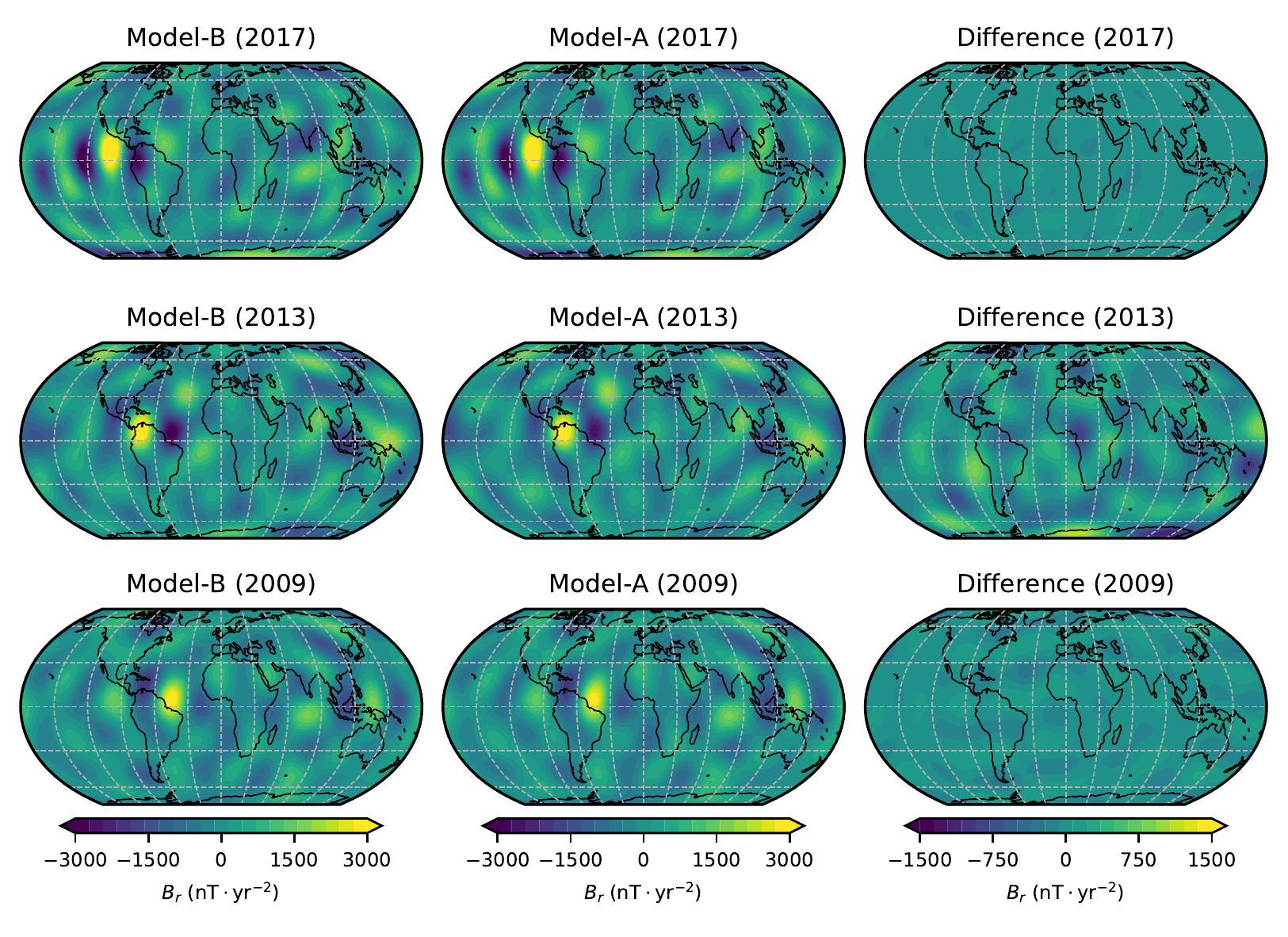}
    \caption{Global maps of the radial SA up to degree 10 at the CMB for \mbox{Model-B} (left column), \mbox{Model-A} (center column) and the difference (right column). The maps are computed in 2009 (bottom row), 2013 (center row) and 2017 (top row). The projection is Equal Earth \cite[]{Savric2018}.}
    \label{fig:SA_surface_maps}
\end{figure*}

To investigate the effect of platform magnetometer data on the recovered SA, we show in Fig.~\ref{fig:SA_maps} time-longitude maps of the radial SA on the Equator at the CMB computed with \mbox{Model-B} (left) and \mbox{Model-A} (center) alongside the difference map (right).
Recall that \mbox{Model-B} is partly based on platform magnetometer data in contrast to \mbox{Model-A}, so that the difference of the two reflects the use of these data. Both models show the SA pulses in 2009, 2013 and most recently in 2017 as enhancement of the radial SA on the Equator. Of special interest is the pulse in 2013, right in between periods of high-quality magnetic data from the CHAMP and \textit{Swarm} missions. In the difference map, the SA during CHAMP and \textit{Swarm} period is largely unchanged, which suggests that the effect of the \mbox{CryoSat-2} and GRACE data is rather minimal during these times. In contrast, the SA in the gap period is distinctly different for the two models. Differences that are large in absolute value seem to be concentrated around \ang{0} and \ang{180} longitude on the Equator which coincides with the Pacific and the region in the South Atlantic close to Central Africa. The geographical location of the differences is more clearly seen in Fig.~\ref{fig:SA_surface_maps}, which shows global maps of the radial SA at the CMB during the SA pulses in 2009, 2013 and 2017.
Again, the difference between \mbox{Model-B} and \mbox{Model-A} is small in 2009 and 2017, i.e.~during CHAMP and \textit{Swarm} times, but large in 2013 in the middle of the gap period. The regions with the largest differences are located in the Southern hemisphere and the Equatorial region with prominent examples in the West and South Pacific Ocean, and Central Africa. Our findings seem to indicate that the platform magnetometers have the desired effect of balancing the uneven spatial distribution of the ground observatory network in the gap period.

\section{Conclusions}
\label{sec:conclusions}

In this study, we present a co-estimation scheme within the framework of the CHAOS field model series that is capable of estimating both a geomagnetic field model and, at the same time, calibration parameters for platform magnetometers. This approach enables us to use platform magnetometer data to supplement high-quality magnetic data from magnetic survey satellites and removes the requirement for utilizing a-priori geomagnetic field models to calibrate platform magnetometer data.

We followed \cite{Alken2020} but went further in that we co-estimated a model of not only the internal field but also the external field. The co-estimation scheme relies on absolute magnetic data which we took from CHAMP, \mbox{Swarm-A}, \mbox{Swarm-B} and the monthly SV data from ground observatories between 2008 and 2018. Magnetic data from five platform magnetometers were used: three on-board \mbox{CryoSat-2} and one on-board each of the GRACE satellite pair. This allowed us to considerably improve the geographical and temporal coverage of satellite data after CHAMP and before the launch of the \textit{Swarm} satellites.

We successfully co-estimated a geomagnetic field model along with calibration parameters of the five platform magnetometers. The misfit to the high-quality satellite data and ground observatory SV data was similar to that for models derived without including platform magnetometer data, and the good fit to an independent ground observatory dataset from Saint Helena provide evidence that our modeling approach performs well.

In a series of experiments we investigated the trade-offs when co-estimating calibration and geomagnetic field model parameters. We found that the calibration parameters strongly correlate with the internal axial dipole and the RC-baseline corrections of the external field during the gap period, when there is less high-quality data available. By preventing platform magnetometer data from contributing to the internal axial dipole and using constant RC-baseline corrections throughout the entire gap period, we successfully avoided those complications.

Our experiments showed that including platform magnetometer data leaves the SA signal practically unchanged during the CHAMP and \textit{Swarm} period but leads to differences in the gap period. The difference in the recovered SA signal is stronger in the West and South Pacific, where only a few observatories are located, which suggests that platform magnetometer data help to improve the global picture of the SA. Based on our investigations, we find that it is worthwhile to include platform magnetometer data in internal field modeling, in particular from CryoSat-2 given the relative low noise level.

\section*{Abbreviations}

\begin{description}
    \item[CHAMP] CHAllenging Minisatellite Payload
    \item[CMB] Core mantle boundary
    \item[CRF] Common reference frame
    \item[DMSP] Defense Meteorological Satellite Program
    \item[FGM] Fluxgate magnetometer
    \item[GRACE] Gravity Recovery and Climate Experiment
    \item[GSM] Geocentric solar magnetic
    \item[NEC] North-east-center
    \item[nT] NanoTesla
    \item[QD] Quasi-dipole
    \item[RTP] Radius-theta-phi
    \item[SM] Solar magnetic
    \item[VFM] Vector field magnetometer
\end{description}

%%%%%%%%%%%%%%%%%%%%%%%%%%%%%%%%%%%%%%%%%%%%%%
%%                                          %%
%% Backmatter begins here                   %%
%%                                          %%
%%%%%%%%%%%%%%%%%%%%%%%%%%%%%%%%%%%%%%%%%%%%%%

\begin{backmatter}

\section*{Declarations}

\subsection*{Availability of data and materials}

The datasets supporting the conclusions of this article are available in the following repositories:\\
\textit{Swarm} and \mbox{CryoSat-2} data are available from \url{https://earth.esa.int/web/guest/swarm/data-access};\\
The GRACE data are available from \url{ftp://ftp.spacecenter.dk/data/magnetic-satellites/GRACE/}\\
CHAMP data are available from \url{https://isdc.gfz-potsdam.de/champ-isdc};\\
Ground observatory data are available from \url{ftp://ftp.nerc-murchison.ac.uk/geomag/Swarm/AUX_OBS/hour/};\\
The RC-index is available from \url{http://www.spacecenter.dk/files/magnetic-models/RC/};\\
The \mbox{CHAOS-6} model and its updates are available from \url{http://www.spacecenter.dk/files/magnetic-models/CHAOS-6/}, and;\\
Solar wind speed, interplanetary magnetic field and Kp-index are available from \url{https://omniweb.gsfc.nasa.gov/ow.html}.

\subsection*{Competing interests}

The authors declare that they have no competing interests.

\subsection*{Funding}

CK and CCF were funded by the European Research Council (ERC) under the European Union’s Horizon 2020 research and innovation programme (grant agreement No. 772561). The study has been partly supported as part of \textit{Swarm} DISC activities, funded by ESA contract no. 4000109587.

\subsection*{Author's contributions}

CK developed the modeling software for the co-estimation of calibration parameters, derived the presented models and led the writing of the manuscript. CCF participated in the design of the study. NiO pre-processed the CryoSat-2 and GRACE platform magnetometer data, developed the CHAOS modeling approach and participated in the design of the study. All co-authors read and approved the final manuscript.

\subsection*{Acknowledgements}

The European Space Agency (ESA) is gratefully acknowledged for providing
access to the \textit{Swarm} L1b data, CryoSat‑2 and GRACE platform magnetometer data and related engineering information. We wish to thank the German Aerospace Center (DLR) and the Federal Ministry of Education and Research for supporting the CHAMP mission. Furthermore, we would like to thank the staff of the geomagnetic observatories and INTERMAGNET for supplying high-quality observatory data. Susan Macmillan (BGS) is gratefully acknowledged for collating checked and corrected observatory hourly mean values in the AUX OBS database.

%%  After submission of the .TEX file,                     %%
%%  you will be prompted to submit your .BBL file.         %%

% if your bibliography is in bibtex format, use those commands:
\bibliographystyle{agu08} % Style BST file (bmc-mathphys, vancouver, spbasic, aps-nameyear).
\bibliography{references}      % Bibliography file (usually '*.bib' )

\end{backmatter}
\end{document}